\pdfoutput=1
\newcommand*{\ATLASLATEXPATH}{}
\documentclass[PAPER, cernpreprint, texlive=2016, USenglish]{\ATLASLATEXPATH atlasdoc}
 
\usepackage{\ATLASLATEXPATH atlaspackage}
\usepackage{\ATLASLATEXPATH atlasbiblatex}
 
\usepackage{\ATLASLATEXPATH atlasphysics}
 
\graphicspath{{logos/}{figures/}}
 
\usepackage{ANA-EXOT-2018-30-PAPER-defs}
 
\usepackage{multirow}

\AtlasTitle{Search for a heavy charged boson in events with a charged
lepton and missing transverse momentum from $pp$ collisions
at $\sqrt{s} = 13$~\TeV\ with the ATLAS detector}

\AtlasAbstract{
A search for a heavy charged-boson resonance decaying into a charged lepton (electron or muon)
and a neutrino is reported. A data sample of 139~fb$^{-1}$ of proton--proton collisions at
$\sqrt{s} = 13$~\TeV\ collected with the ATLAS detector at the LHC during 2015--2018 is used in the search.
The observed transverse mass distribution computed from the lepton and missing transverse
momenta is consistent with the distribution expected from the Standard Model, and upper limits
on the cross section for $pp \to W^\prime \to \ell\nu$ are extracted ($\ell = e$ or $\mu$).
These vary between 1.3~pb and 0.05~fb depending on the
resonance mass in the range between 0.15 and 7.0~\TeV\ at 95\% confidence level for the
electron and muon channels combined. Gauge bosons with a mass below 6.0~\TeV\ and 5.1~\TeV\
are excluded in the electron and muon channels, respectively, in a model with a resonance
that has couplings to fermions identical to those of the Standard Model $W$ boson.
Cross-section limits are also provided for resonances with several fixed $\Gamma / m$
values in the range between 1\% and 15\%. Model-independent limits are derived in single-bin
signal regions defined by a varying minimum transverse mass threshold.
The resulting visible cross-section upper limits range between 4.6 (15)~pb and 22 (22)~ab as
the threshold increases from 130 (110)~\GeV\
to 5.1 (5.1)~\TeV\ in the electron (muon) channel.}
 
\author{The ATLAS Collaboration}
 
\AtlasRefCode{EXOT-2018-30}
 
\PreprintIdNumber{CERN-EP-2019-100}
 
\AtlasJournalRef{Phys. Rev. D 100 (2019) 052013}
 
\AtlasDOI{10.1103/PhysRevD.100.052013}

\AtlasCoverSupportingNote{Supporting documentation}{https://cds.cern.ch/record/2640567}
\AtlasCoverSupportingNote{Analysis updates since draft 1}{https://cds.cern.ch/record/2640567/files/draft2updates.pdf}
 
\AtlasCoverCommentsDeadline{16 May 2019}
 
\AtlasCoverAnalysisTeam{M.~K.~Bugge, J.~Catmore, K.~Choi,
D.~Hayden, U.~Klein, M.~W.~O'Keefe,
F.~Ould-Saada, S.~Willocq}
 
\AtlasCoverEdBoardMember{S.~Angelidakis, F.~Ledroit-Guillon~(chair), C.~Lee}
 
\AtlasCoverEgroupEditors{atlas-EXOT-2018-30-editors@cern.ch}
 
\AtlasCoverEgroupEdBoard{atlas-EXOT-2018-30-editorial-board@cern.ch}

\hypersetup{pdftitle={ATLAS document},pdfauthor={The ATLAS Collaboration}}
 
\begin{document}
 
\maketitle

\section{Introduction}
\newcommand{\AtlasCoordFootnote}{
ATLAS uses a right-handed coordinate system with its origin at the nominal interaction point (IP)
in the center of the detector and the $z$-axis along the beam pipe.
The $x$-axis points from the IP to the center of the LHC ring,
and the $y$-axis points upwards.
Cylindrical coordinates $(r,\phi)$ are used in the transverse plane,
$\phi$ being the azimuthal angle around the $z$-axis.
The pseudorapidity is defined in terms of the polar angle $\theta$ as $\eta = -\ln \tan(\theta/2)$.
Angular distance is measured in units of $\Delta R \equiv \sqrt{(\Delta\eta)^{2} + (\Delta\phi)^{2}}$.}

One of the main goals of the Large Hadron Collider (LHC) remains the search for
physics beyond the Standard Model (SM). Much progress has been made in this search
thanks to a broad program that encompasses many different final states.
Leptonic final states provide a low-background and efficient experimental signature that
brings excellent sensitivity to new phenomena at the LHC.
In this article, the results of a search for resonances decaying into a charged lepton
and a neutrino are presented, based on \totlumi~\ifb\ of proton--proton ($pp$) collisions
at a center-of-mass energy of 13~\TeV.
The data were collected with the ATLAS detector during the 2015--2018 running period of the LHC,
referred to as Run 2.
 
The search results are interpreted in terms of the production of a heavy spin-1 \wp\ boson
with subsequent decay into the $\ell\nu$ final state ($\ell = e$ or $\mu$).
Such production is predicted in many models of physics beyond the SM
as in Grand Unified Theory models,
Left--Right symmetry models~\cite{Mohapatra:1974hk,Senjanovic:1975rk},
Little Higgs models~\cite{ArkaniHamed:2002qy},
or models with extra dimensions~\cite{Appelquist:2000nn,Agashe:2008jb},
most of which aim to solve the hierarchy problem.
The interpretation in this article uses a simplified model, referred to as the
Sequential Standard Model (SSM)~\cite{ssm}, in which the \wp\ boson couples to fermions with the
same strength as the $W$ boson in the SM but with suppressed coupling to SM bosons.
Alternative interpretations in terms of generic resonances with different fixed widths
($\Gamma/m$ between 1\% and 15\%) are also provided
for possible reinterpretation in the context of other models.
Finally, results are also presented in terms of model-independent upper limits on the
number of signal events and on the visible cross section.
 
Previous searches for \wp\ bosons have been carried out at the LHC in leptonic, semileptonic,
and hadronic final states by the ATLAS and CMS collaborations.
The most sensitive searches for \wp\ bosons
are those in the $e\nu$ and $\mu\nu$ channels~\cite{EXOT-2016-06, CMS-EXO-16-033},
with the most stringent limits to date being set by ATLAS and CMS in the analysis of about
36~\ifb\ of $pp$ collisions at $\rts = 13$~\TeV. A lower limit of 5.2~\TeV\ is set on the \wp\
boson mass in the electron channel~\cite{EXOT-2016-06} and 4.9~\TeV\
in the muon channel~\cite{CMS-EXO-16-033}, at the 95\% confidence level (CL) in the SSM.
 
The search relies on events collected using single-electron or single-muon triggers
with high transverse momentum thresholds.
The dominant background source originates from Drell--Yan (DY) production of $W$~bosons.
Discrimination between signal and background events relies on the
transverse mass $(\mt)$ computed from the charged-lepton transverse momentum $(\pt)$ and the
missing transverse momentum (whose magnitude is denoted \met) in the event:
\begin{equation*}
\mt = \sqrt{2\,\pt\,\met{}\,(1-\cos\phi_{\ell\nu})}\: ,
\end{equation*}
 
where $\phi_{\ell\nu}$ is the angle between the charged lepton and missing transverse
momentum directions in the transverse plane.\footnote{\AtlasCoordFootnote}
Final interpreted results are based on a statistical analysis in which the
shape of the signal and both the shape and normalization of the background expectations are derived from
Monte Carlo (MC) simulation, except for the background contribution arising from jets misidentified
as leptons or from hadron decays.
The results presented in this article, compared with those from Ref.~\cite{EXOT-2016-06},
benefit from: an increase in the integrated luminosity by a factor of four;
several upgrades in reconstruction software, including a new algorithm for electron
reconstruction~\cite{EGAM-2018-01} and an improved treatment of the relative alignment
between the inner tracker and the muon spectrometer; and several interpretations with reduced
or no model dependence.

\section{ATLAS detector}
 
The ATLAS experiment~\cite{PERF-2007-01} at the LHC is a multipurpose particle detector
with a forward--backward symmetric cylindrical geometry and a near $4\pi$ coverage in
solid angle.
It consists of an inner detector for tracking surrounded by a thin superconducting solenoid
providing a \SI{2}{\tesla} axial magnetic field, electromagnetic (EM) and hadronic calorimeters, and a muon spectrometer.
The inner detector covers the pseudorapidity range $|\eta| < 2.5$.
It consists of silicon pixel, silicon microstrip, and transition radiation tracking detectors.
An additional innermost pixel layer~\cite{ATLAS-TDR-19,Abbott:2018ikt} inserted at a radius of 3.3~cm has been used since 2015.
Liquid-argon (LAr) sampling calorimeters provide EM energy measurements
with high granularity.
A hadronic scintillator-tile calorimeter covers the central pseudorapidity range ($|\eta| < 1.7$).
The endcap and forward regions are instrumented with LAr calorimeters
for both the EM and hadronic energy measurements up to $|\eta| = 4.9$.
The muon spectrometer surrounds the calorimeters and features
three large air-core toroidal superconducting magnet systems with eight coils each.
The field integral of the toroids ranges between \num{2.0} and \SI{6.0}~Tm
across most of the detector. The muon spectrometer includes a system of precision tracking
chambers and fast detectors for triggering.
A two-level trigger system~\cite{TRIG-2016-01} is used to select events.
The first-level trigger is implemented in hardware and uses a subset of the detector information
to reduce the accepted rate to at most \SI{100}{\kilo\hertz}.
This is followed by a software-based trigger level that reduces the accepted event rate
to \SI{1}{\kilo\hertz} on average.

\section{Data and Monte Carlo simulation samples}
 
The data for the analysis were collected during Run 2 at the LHC at $\rts = 13$~\TeV\
and correspond to an integrated luminosity of \totlumi~\ifb\ after the requirement that
beams were stable, all detector systems were functional, and the data satisfied a set
of quality criteria.
Single-electron triggers required that electron candidates satisfy either medium identification
criteria~\cite{EGAM-2018-01}
and have a transverse energy $\ET > 60$~\GeV\ or loose identification criteria and have $\ET > 140$~\GeV.
For the 3.2~\ifb\ collected in 2015, the \ET\ thresholds were 24 and 120~\GeV, respectively.
Single-muon triggers required the presence of at least one muon reconstructed in both the inner
detector and the muon spectrometer with $\pt > 50$~\GeV.
The trigger efficiency for DY $W$~boson events (relative to the full event selection described in Section~\ref{sec:selection})
is estimated to be 99\% in the electron channel and 85\% in the muon channel,
with little dependence on the \mt\ value.
 
\label{sec:simulation}
 
Signal MC events with $\wp \to e\nu$ and $\wp \to \mu\nu$ decays in the SSM were produced
at leading order (LO) with the \PYTHIAV{v8.183} event generator~\cite{pythia8}
and the NNPDF23LO parton distribution function (PDF) set~\cite{Ball:2012cx}.
The A14 set of tuned parameters (i.e.\ the A14 tune)~\cite{ATL-PHYS-PUB-2014-021} was used for
the parton showering and hadronization process.
In the SSM, the couplings of the \wp~boson to SM fermions are chosen to be identical to those of the SM
$W$~boson, whereas the couplings to SM bosons are set to zero.
The corresponding branching fraction for \wp~boson decays into leptons of one generation
is 10.8\% for $m(\wp) = 150$~\GeV\ and decreases above the $tb$ threshold to
a nearly constant value of 8.2\% for $m(\wp)$ above 1~\TeV.
Similarly, the ratio of the \wp~boson width to its mass varies from 2.7\% for $m(\wp) = 150$~\GeV\
to 3.5\% above the $tb$ threshold.
Decays into the $\tau\nu$ final state with subsequent leptonic decay of the $\tau$-lepton are not
included as they were found to add negligible signal acceptance in previous studies~\cite{EXOT-2012-02}.
Interference between \wp\ and $W$ boson production is not included in this analysis.
 
The dominant background due to DY production of $W$~bosons decaying
into $e\nu$, $\mu\nu$, and $\tau\nu$ final states was simulated at next-to-leading order (NLO)
with the \POWHEGBOX\ v2 event generator~\cite{Nason:2004rx,Frixione:2007vw,Alioli:2008gx,Alioli:2010xd} using the
CT10 PDF set~\cite{ct10}.
Background events from DY production of $Z/\gamma^*$~bosons decaying into $ee$, $\mu\mu$, and $\tau\tau$
final states were also simulated with the same event generator and PDF set.
In both cases, \PYTHIAV{v8.186} was used for the parton showering and hadronization process
with the AZNLO tune~\cite{STDM-2012-23}.
The DY processes were generated separately in different $\ell\nu$ or $\ell\ell$ mass ranges to
guarantee that sufficiently large numbers of events remain after event selection in the full mass range
relevant to the analysis.
Cross sections calculated by \POWHEGBOX\ for both DY processes were corrected via mass-dependent
$K$-factors to account for
QCD effects at next-to-next-to-leading order (NNLO) and electroweak (EW) effects at NLO.
The QCD corrections were computed with {\textsc{VRAP}} v0.9~\cite{vrap} and
the CT14 NNLO PDF set~\cite{CT14}. These corrections increased the cross section by about 5\% for
$m_{\ell\nu} = 1$~\TeV\ and 15\% for $m_{\ell\nu} = 6$~\TeV. The EW corrections were computed with
{\textsc{MCSANC}}~\cite{MCSANC} in the case of QED effects due to initial-state radiation,
interference between initial- and final-state radiation, and Sudakov logarithm single-loop corrections.
These corrections were added to the NNLO QCD cross-section prediction in the so-called additive approach
(see Section~\ref{sec:systematics}) because of a lack of calculations of mixed QCD and EW terms.
As a result, the cross section decreased by about 10\% for $m_{\ell\nu} = 1$~\TeV\ and 20\% for
$m_{\ell\nu} = 6$~\TeV.
The effects due to QED final-state radiation were already included in the event generation
using {\textsc{PHOTOS++}}~\cite{photos}.
The QCD corrections based on {\textsc{VRAP}} and the CT14 NNLO PDF set were also applied to the
signal samples. No electroweak corrections, beyond those already accounted for with {\textsc{PHOTOS++}},
were applied to the signal samples as those are model-dependent.
 
Additional background sources from diboson ($WW$, $WZ$, and $ZZ$) production were simulated with
the \SHERPA~v2.2.1 event generator~\cite{Gleisberg:2008ta} and the NNPDF30 NNLO PDF set~\cite{NNPDF30}.
These processes were computed at NLO for up to one additional parton and at LO for up to three partons.
The production of top-quark pairs and single top quarks (in the $s$- and $Wt$-channels)
was performed at NLO with \POWHEGBOX~\cite{Frixione:2007nw,Alioli:2009je,Re:2010bp} and the NNPDF30 NLO PDF set, interfaced with
\PYTHIAV{v8.183} and the A14 tune.
Single top-quark production in the $t$-channel was performed in the same way except for the use
of the NNPDF3.04f NLO PDF set.
The cross sections used to normalize the diboson MC samples are computed with \SHERPA,
and the top-quark pair cross section is taken to be $832^{+46}_{-52}$~pb for a top-quark
mass of 172.5~\GeV. This value is calculated at NNLO in QCD, including
the summation of next-to-next-to-leading logarithmic soft gluon terms, with
Top++2.0~\cite{Cacciari:2011hy,Beneke:2011mq,Baernreuther:2012ws,Czakon:2012zr,Czakon:2012pz,Czakon:2013goa,Czakon:2011xx}.
A correction depending on the top-quark \pt\ value is applied to account for shape effects due
to NNLO QCD and NLO EW corrections according to Ref.~\cite{Czakon:2017wor}.
The cross sections for single top-quark production are computed at approximate NNLO
accuracy~\cite{t-channel,s-channel,wt}.
 
For all MC samples, except those produced with \SHERPA,
$b$-hadron and $c$-hadron decays were handled by {\textsc{EvtGen}} v1.2.0~\cite{Lange:2001uf}.
Inelastic $pp$ events generated using \PYTHIAV{v8.186} with the A3 tune~\cite{ATL-PHYS-PUB-2016-017}
and the NNPDF23LO PDF set
were added to the hard-scattering interaction in such a way as to reproduce the effects of additional
$pp$ interactions in each bunch crossing during data collection (pileup).
The detector response was simulated with \GEANT~4~\cite{geant, SOFT-2010-01},
and the events were processed with the same reconstruction software as for the data.
Energy/momentum scale and efficiency corrections are applied to the results of the simulation to account
for small differences between the simulation and the performance measured directly from the
data~\cite{EGAM-2018-01, PERF-2015-10}.
 

\section{Event reconstruction and selection}
 
The analysis relies on the reconstruction and identification of electrons and muons,
as well as the missing transverse momentum in each event.
Collision vertices are reconstructed with inner detector tracks that satisfy $\pt > 0.5$~\GeV,
and the primary vertex is chosen as the vertex with the largest $\Sigma \pt^2$ for the tracks associated with the vertex.
 
Electron candidates are reconstructed by matching inner detector tracks to clusters of
energy deposited in the EM calorimeter.
Electrons must lie within $|\eta| < 2.47$, excluding the barrel--endcap
transition region defined by $1.37 < |\eta| < 1.52$, and satisfy calorimeter energy cluster
quality criteria. The cluster must have $\ET > 65$~\GeV, and
the associated track must have a transverse
impact parameter significance relative to the beam axis $|d_0|/\sigma_{d_0} < 5$.
Successful candidates are identified with a likelihood method and need to satisfy the
tight identification criteria~\cite{EGAM-2018-01}. The likelihood relies on
the shape of the EM shower measured in the calorimeter, the quality of the track
reconstruction, and the quality of the match between the track and the cluster.
To suppress electron candidates originating from photon conversions, hadron decays, or
jets misidentified as electrons (hereafter referred to as fake electrons),
electron candidates are required to satisfy the gradient
isolation criteria~\cite{EGAM-2018-01} based on both tracking and calorimeter measurements.
The reconstruction and identification efficiency rises from approximately 80\% at $\pt = 60$~\GeV\ to 90\% above
500~\GeV, and the isolation efficiency is slightly higher than 99\% for \pt\ values above 200~\GeV.
The electron energy resolution for $\et > 1$~\TeV\ can be characterized by $\sigma(E)/E = c_e$,
with $c_e$ varying between 0.007 and 0.012~\cite{EGAM-2018-01}
in the range $|\eta| < 1.2$ which dominates the high-mass part of the search.
The corresponding \mt\ resolution ranges from approximately 1.3\% at \mt\ values
near 2~\TeV\ to 1.0\% near 6~\TeV.
 
Muon candidates are reconstructed by matching inner detector tracks with muon spectrometer tracks
and by reconstructing a final track combining the measurements from both detector systems
while taking the energy loss in the calorimeter into account.
The candidates must satisfy quality selection criteria optimized for high-\pt\ performance~\cite{PERF-2015-10}
by requiring the candidate tracks to have associated measurements in the three different chamber layers
of the muon spectrometer. The tracks must also have consistent charge-to-momentum ratio measurements in
the inner detector and muon spectrometer,
have sufficiently small relative uncertainty in the charge-to-momentum ratios for the combined tracks,
and be located in detector regions with high-quality chamber alignment.
Candidates must have $|\eta| < 2.5$, $\pt > 55$~\GeV, $|d_0|/\sigma_{d_0} < 3$, and
$|z_0| \sin\theta < 0.5$~mm, where $z_0$ is the longitudinal impact parameter relative to
the primary vertex.
The reconstruction and identification efficiency is 69\% for $\pt = 1$~\TeV\ and
decreases to 57\% for $\pt = 2.5$~\TeV.
Muon candidates from hadron decays are suppressed by imposing a track-based isolation~\cite{PERF-2015-10}
that achieves an efficiency higher than 99\% for the full \pt\ range of interest.
The muon \pt\ resolution at $\pt > 1$~\TeV\ can be described as $\sigma(\pt)/\pt = c_\mu\:\pt$,
with $c_\mu$ varying between 0.08 and 0.20~\TeV$^{-1}$ depending on the detector region~\cite{PERF-2015-10}.
This resolution dominates the \mt\ resolution in the muon channel.
 
Jets are reconstructed from topological clusters of energy deposits in calorimeter cells~\cite{PERF-2014-07}
with the \antikt\ clustering algorithm~\cite{Cacciari:2008gp} implemented in FastJet~\cite{Cacciari:2011ma}.
A radius parameter $R$ equal to 0.4 is used, and the clusters are calibrated at the
EM scale~\cite{PERF-2016-04}.
Jets are required to have $\pt > 20$ (30)~\GeV\ for $|\eta|$ smaller (greater) than 2.4.
To remove jets originating from pileup, jet-vertex-tagging is applied~\cite{PERF-2014-03}.
 
The event's missing transverse momentum is computed as the
vectorial sum of the transverse momenta of leptons, photons, and jets.
The overlap between these is resolved according to Ref.~\cite{PERF-2016-07}.
Electrons and muons must pass the selection requirements described above.
In addition to the above particles and jets, the \met\ calculation includes a \textit{soft term}~\cite{PERF-2016-07}
accounting for the contribution from tracks associated with the primary vertex but not
associated with leptons, converted photons, or jets already included in the \met\ calculation.
 
 
\label{sec:selection}
 
Events are required to have a primary vertex.
They are rejected if any of the jets fail to pass a cleaning procedure designed to
suppress non-collision background and calorimeter noise~\cite{ATLAS-CONF-2015-029}.
 
In the electron channel, events must have exactly one electron passing the selection
described above. 
Events are vetoed if they contain any additional electron candidate satisfying the medium selection criteria
and having $\pt > 20$~\GeV. Events are also vetoed if they contain any muon candidate satisfying the medium
selection criteria and having $\pt > 20$~\GeV. The missing transverse momentum must satisfy $\met > 65$~\GeV,
and the transverse mass must satisfy $\mt > 130$~\GeV.
In the muon channel, events must have exactly one selected muon as detailed above,
and the same veto on additional electron and muon candidates
is applied, except that electron candidates close to the muon ($\Delta R < 0.1$) are assumed to
arise from photon radiation from the muon and are thus not considered as additional electron candidates.
Events are required to satisfy $\met > 55$~\GeV\ and $\mt > 110$~\GeV\ in the muon channel.
The event selection described above defines the signal regions in the electron and muon channels.
In these regions, the acceptance times efficiency
for \wp\ signal events decreases from 79\% (52\%) to 64\% (44\%)
as the \wp\ boson mass increases from 2 to 7~\TeV\ in the electron (muon) channel.
The decrease at high $m(\wp)$ is generally due to the combined effect of a growing low-mass tail
at larger $m(\wp)$ and the kinematic selection thresholds. In the case of the muon channel,
it also originates from a decrease in the identification efficiency at
higher \pt\ values due to the requirements on the charge-to-momentum measurement.

\section{Background estimation and event yields}
 
The background from DY production of $W$ and $Z/\gamma^*$ bosons as well as from top-quark pair,
single top-quark, and diboson production is modeled with the MC samples described in
Section~\ref{sec:simulation}.
To compensate for the limited number of events at high \mt, the smoothly falling
\mt\ distributions for top-quark (corresponding to both pair and single production) and diboson
samples are fitted and extrapolated to high \mt\ with the following functions commonly used in dijet searches (e.g.\ Refs.~\cite{EXOT-2016-21, CMS-EXO-16-056}):
 
\begin{align}
f^\textrm{bkg1}(\mt) = \textrm{e}^{-a} \mt^{b}\, \mt^{c \log(\mt)}, ~\textrm{and}
\nonumber  
\end{align}
\begin{align}
f^\textrm{bkg2}(\mt) = \frac{a}{(\mt+b)^{c}}.
\label{eq:powerlaw}
\end{align}
Function $f^\textrm{bkg1}$ is the nominal extrapolation function for the top-quark background
in both the electron and muon channels as well as for the diboson background in the electron channel.
Function $f^\textrm{bkg2}$ is the nominal function for the diboson background in the muon channel.
In all cases, checks are performed to guarantee that the function reproduces the
event yields at lower~\mt\ values and that its cumulative distribution (starting from the highest \mt\
values) is consistent with the small integrated event yields available in the MC samples.
 
The background contribution from events with fake electrons or muons mostly originates
from multijet production and is extracted
from the data using the same matrix method as used in previous analyses and described
in Ref.~\cite{EXOT-2015-06}.
This method relies on data samples in which the electron or muon selection is loosened (referred to as the \textit{loose} selection).
The efficiency for those lepton candidates to pass the nominal lepton selection (\textit{tight}) is measured to derive
an estimate of the background from fake leptons. The \textit{loose} selection is close to that applied
by the trigger requirements. 
The fraction $f$ of fake leptons passing the \textit{loose} selection that also pass the nominal lepton selection is estimated from the data
in background-enriched control regions that are orthogonal to the signal regions.
These control regions are built by requiring that there are no $Z \to \ell\ell$ candidates formed by combining the selected lepton with a loose lepton in the event and that the \met\ value is less than 60 (55)~\GeV\ in the electron (muon) channel. Additional requirements are placed on the minimum impact parameter, the presence of at least one jet, and the proximity of the missing transverse momentum vector to the lepton in the muon channel to reduce the contribution from prompt muons.
The remaining contributions from prompt electrons and muons in these control regions are subtracted using MC simulation.
The number of jets misidentified as leptons ($N_\textrm{T}^\textrm{multijet}$) in the
signal regions is computed as
 
\begin{align*}
N_\textrm{T}^\textrm{multijet} = f \, N_\textrm{F} = \frac{f}{r-f} \left[r\, (N_\textrm{L}+N_\textrm{T}) - N_\textrm{T} \right],
\end{align*}
 
where $N_\textrm{F}$ is the number of fake leptons that pass the \textit{loose} lepton selection, $N_\textrm{L}$ is the number of lepton candidates that
pass the \textit{loose} lepton selection but fail the nominal lepton selection, and $N_\textrm{T}$ is the number of lepton
candidates that pass the nominal lepton selection. The numbers $N_\textrm{L}$ and $N_\textrm{T}$ are extracted
from the signal regions.
In addition, the quantity $r$, corresponding to the fraction of real leptons satisfying
the nominal selection in the sample of \textit{loose} candidates, is computed from the DY $W$ boson MC samples.
Like for the top-quark and diboson background sources, the \mt\ distribution is extrapolated to high
values by using a function with the same form as in Eq.~(\ref{eq:powerlaw}) in the electron
channel and the function $f^\textrm{multijet}(\mt) = a\mt^{-b}$ in the muon channel.
The same set of checks concerning the quality of the extrapolation are performed as for the top-quark
and diboson backgrounds.

The \mt\ distributions in data and simulation are shown in Figure~\ref{fig:mt_panels},
and the numbers of events in several \mt\ ranges are presented in Table~\ref{tab:mt_bins}.
No event is observed beyond \mt\ values of 10~\TeV\ in either channel.
The features observed in these distributions are discussed in Section~\ref{sec:results}.
The DY $W$ boson contribution dominates the total background with a fraction varying
between approximately 69\% (72\%) and 95\% (88\%) in the electron (muon) channel.
Other background contributions arise mostly from DY $Z/\gamma^*$ boson, top-quark, and diboson production.
The contribution from multijet events in the electron channel decreases from approximately 10\%
at the lowest \mt\ values to less than 5\% at high \mt, and in the muon channel it is less
than 3.2\% (1.7\%) for \mt\ values below (above) 600~\GeV.
 
\begin{figure}[htbp]
\centering
\includegraphics[width=0.58\textwidth]{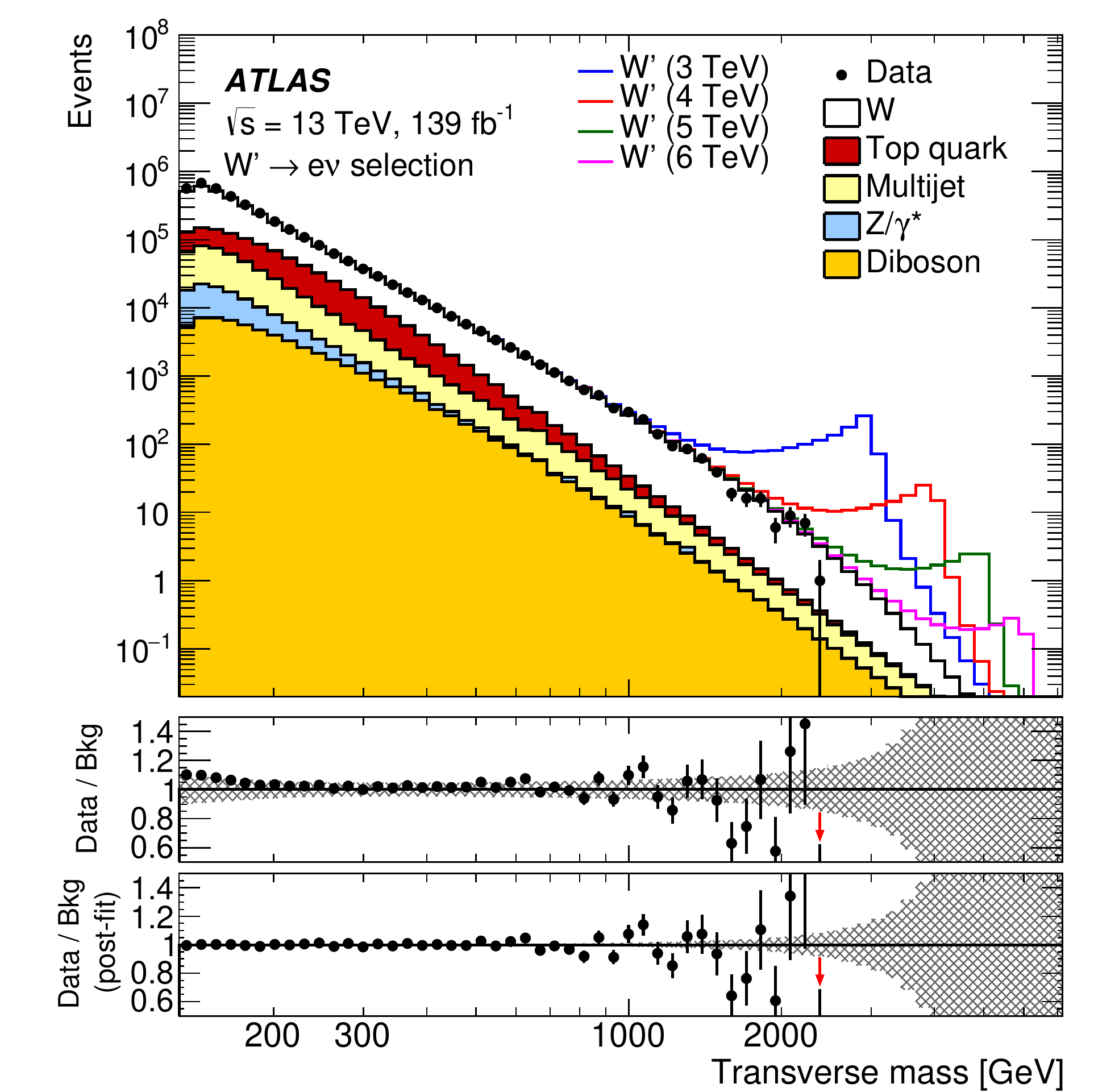}
\includegraphics[width=0.58\textwidth]{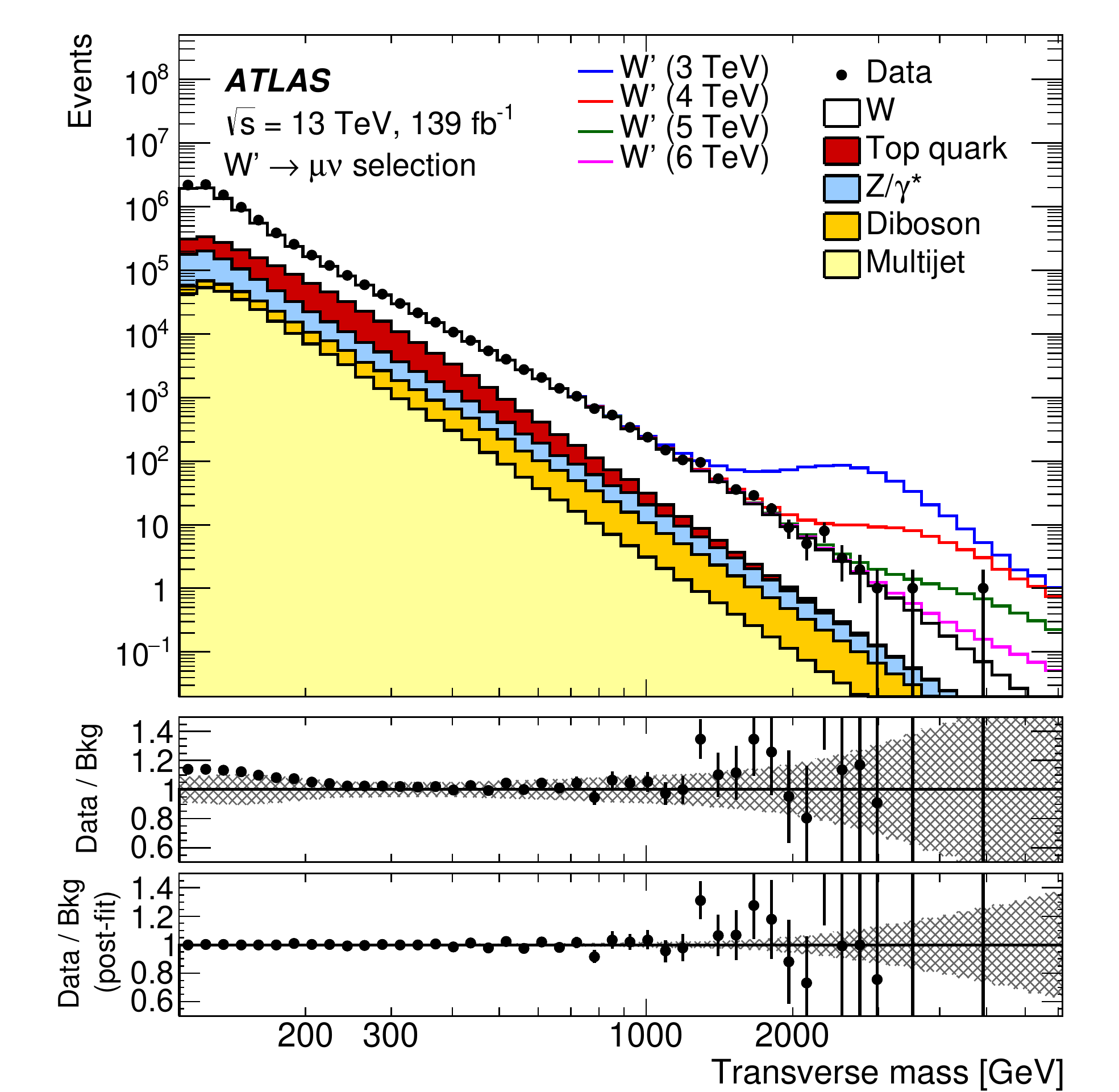}
\caption{Distributions of the transverse mass for data and predicted background events in the
electron (top) and muon (bottom) channels.
Expected signal distributions for several SSM \wp\ boson masses are shown stacked on top of the
total expected background. The middle panels show ratios of the number of events observed in the data to the
expected total background count, while the lower panels show the same ratio when
taking into account the pulls on the nuisance parameters observed in the statistical analysis
(Section~\ref{sec:results}). The hatched bands represent the total uncertainty in the background estimate (Section~\ref{sec:systematics}).
Arrows in the middle and lower panels for the electron channel indicate data points that lie outside the vertical axis range.
}
\label{fig:mt_panels}
\end{figure}
 
\begin{table}[tb]
\caption{
Number of events in the data and the total expected background passing the full event
selection in different \mt\ ranges.
Expected numbers of \wp\ signal events are provided for several different masses.
The uncertainties include both statistical and systematic sources of uncertainty.}
\footnotesize
\centering
\begin{tabular}{l r@{\,$\pm$\,}l r@{\,$\pm$\,}l r@{\,$\pm$\,}l r@{\,$\pm$\,}l r@{\,$\pm$\,}l r@{\,$\pm$\,}l }
\hline\hline
\multicolumn{13}{l}{Electron channel} \\ \hline
\mt\ [\GeV] & \multicolumn{2}{c}{~~130--400} & \multicolumn{2}{c}{~~400--600} & \multicolumn{2}{c}{~~600--1000} & \multicolumn{2}{c}{~~1000--2000} & \multicolumn{2}{c}{~2000--3000} & \multicolumn{2}{c}{3000--10\,000} \\ \hline
Data        & \multicolumn{2}{l}{3\,538\,403} & \multicolumn{2}{l}{35\,568} & \multicolumn{2}{l}{7358} & \multicolumn{2}{l}{~~~~818} & \multicolumn{2}{l}{~17} & \multicolumn{2}{l}{~~0}  \\ 
Background  & 3\,320\,000&250\,000 & 34\,800&1500 & 7200&400 & 830&80 & 20.2&3.1 & 1.3&0.5\\ \hline
\wp\ (2 \TeV)  & 574&22 & 720&40 & 2190&120 & 12200&600 & 1130&290 & 3.20&0.25\\
\wp\ (3 \TeV)  & 68.4&1.9 & 58.6& 2.6 & 127&7 & 448&22 & 860&40 & 87&23\\
\wp\ (4 \TeV)  & 19.6&0.5 & 13.2& 0.5 & 22.1&1.1 & 44.3&2.2 & 49.2&2.3 & 86&4\\
\wp\ (5 \TeV)  & 7.85&0.19 & 4.99& 0.18 & 7.26&0.35 & 9.9&0.5 & 5.82&0.28 & 13.6&0.7\\
\wp\ (6 \TeV)  & 3.76&0.09 & 2.35&0.08 & 3.28&0.16 & 3.82&0.18 & 1.41&0.07 & 2.01&0.10\\ \hline\hline
\multicolumn{13}{l}{Muon channel} \\ \hline
\mt\ [\GeV] & \multicolumn{2}{c}{~~110--400} & \multicolumn{2}{c}{~~400--600} & \multicolumn{2}{c}{~~600--1000} & \multicolumn{2}{c}{~~1000--2000} & \multicolumn{2}{c}{~2000--3000} & \multicolumn{2}{c}{3000--10\,000} \\ \hline
Data & \multicolumn{2}{l}{8\,751\,095} & \multicolumn{2}{l}{26\,225} & \multicolumn{2}{l}{5393} & \multicolumn{2}{l}{~~~~622} & \multicolumn{2}{l}{~~~\,22} & \multicolumn{2}{l}{~~2} \\ 
Background & 7\,800\,000 & 700\,000 & 25\,800 & 1400 & 5300 & 400 & 570 & 50 & 18 & 4 & 2.3 & 0.9 \\ \hline
$W^\prime$ (2 \TeV) & 490 & 14 & 594 & 26 & 1680 & 90 & 6700 & 500 & 1520 & 210 & 70 & 50 \\
$W^\prime$ (3 \TeV) & 58.1 & 1.4 & 45.5 & 1.9 & 102 & 6 & 322 & 31 & 380 & 50 & 160 & 40 \\
$W^\prime$ (4 \TeV) & 16.3 & 0.4 & 9.64 & 0.34 & 15.9 & 0.8 & 32.2 & 3.4 & 34 & 5 & 44 & 13 \\
$W^\prime$ (5 \TeV) & 6.50 & 0.15 & 3.55 & 0.12 & 4.98 & 0.22 & 6.7 & 0.6 & 3.9 & 0.6 & 7.2 & 2.3 \\
$W^\prime$ (6 \TeV) & 3.11 & 0.07 & 1.67 & 0.06 & 2.22 & 0.10 & 2.45 & 0.17 & 0.88 & 0.12 & 1.09 & 0.35 \\ \hline\hline
\end{tabular}
\label{tab:mt_bins}
\end{table}

\section{Systematic uncertainties}
\label{sec:systematics}
 
Systematic uncertainties arise from experimental sources affecting the lepton reconstruction
and identification as well as the missing transverse momentum,
from the data-driven multijet background estimate,
from theoretical sources affecting the shape and normalization of background processes,
and from the extrapolation of background estimates to high \mt\ values.
 
Experimental uncertainties in the electron trigger, reconstruction, identification, and
isolation efficiencies are extracted individually from studies of $Z \to ee$ and $J/\psi \to ee$
decays in the data using a tag-and-probe method~\cite{EGAM-2018-01}.
These studies also yield uncertainties in the electron energy scale and resolution~\cite{EGAM-2018-01}.
Uncertainties in the muon trigger, reconstruction, identification, and isolation efficiencies are
derived from studies of $Z \to \mu\mu$ and $J/\psi \to \mu\mu$ decays in the
data~\cite{PERF-2015-10}.
The muon momentum scale and resolution uncertainties are extracted from those studies as well
as from special chamber-alignment datasets with the toroidal magnetic field turned off~\cite{PERF-2015-10}.
Extrapolation uncertainties toward higher \pt\ are based on the above studies as well as on the simulation.
The impact of those uncertainties is generally small due to the limited \pt\ dependence of the efficiencies,
except for the high-\pt\ muon reconstruction and identification efficiency.
The latter is estimated from differences between data and simulation in the fraction of muons
passing the requirement on the maximum allowed relative error
in the charge-to-momentum ratio measurement. This uncertainty grows with the muon \pt\ up to 35\% (55\%)
for $|\eta| < 1.05$ ($> 1.05$) at the highest \mt\ values probed in this analysis; it becomes a dominant
source of uncertainty at the highest \mt\ values.
Uncertainties in the reconstruction and calibration of jets are taken into account since
those are input to the \met\ calculation.
Finally, all uncertainties affecting electrons, muons, jets, and the soft term are propagated to the
\met\ calculation.
The jet energy resolution and soft term contributions have the largest impact at low \mt,
and their uncertainties are treated as fully correlated between the electron and muon channels.
Uncertainties in the simulation of pileup contributions have little impact on
the \mt\ distribution and are thus neglected.
 
The uncertainty in the multijet background estimate includes the effect of varying the criteria
used in the background-enriched sample selection, and changes in the fractions $f$ are propagated.
As this background estimate is extrapolated with a functional fit at high \mt\ values,
the uncertainty includes the additional impact of variations in the fit range.
In the electron channel, the uncertainty also includes a contribution from the variation
of the functional form due to the larger multijet contribution at high \mt\ in this channel.
This extrapolation uncertainty dominates the overall background uncertainty at \mt\ values
above 3~\TeV\ in the electron channel.
 
No theory uncertainty is applied to the signal.
Uncertainties in the theory inputs used for the background estimation are evaluated as follows.
One of the largest uncertainties affecting the dominant DY background
comes from the use of 90\% CL eigenvector variations for the CT14 NNLO PDF set.
This uncertainty range encompasses the predictions based on the ABM12~\cite{ABM12lhc},
CT10~\cite{ct10}, MMHT14~\cite{mmht2014}, and JR14~\cite{JR14} PDF sets.
It also allows for a sufficiently robust range of predictions in the very high mass region
(i.e.\ at high Bjorken-$x$). In addition, a reduced set of CT14 NNLO PDF eigenvectors that preserves
the potential mass-dependent shape changes is used in the limit-setting procedure.
The PDF uncertainty is enlarged in specific $\ell\nu$ mass regions to encompass
the DY prediction based on the alternative NNPDF30 PDF set if this prediction lies outside the
range from the CT14 NNLO eigenvector variations.
A smaller PDF choice uncertainty is obtained in the muon channel at high \mt\ values than
in the electron channel because the significantly worse muon \pt\ resolution causes migration of events
from low \mt\ values (where the PDF uncertainty is small) to high \mt\ values.
The uncertainty in the mass-dependent $K$-factors used to correct the mass distributions to
predictions at NNLO accuracy in $\alpha_\textrm{s}$ is evaluated by simultaneously varying the renormalization and
factorization scales up and down by factors of two. The largest change (up or down) at each
mass value is then applied as a symmetric scale uncertainty.
The EW correction uncertainty is taken to be the difference between the predictions obtained
with either the multiplicative scheme $((1+\delta_\textrm{EW})\times(1+\delta_\textrm{QCD}))$
or the additive scheme $(1+\delta_\textrm{EW}+\delta_\textrm{QCD})$ for the combination of
higher-order EW ($\delta_\textrm{EW}$) and QCD ($\delta_\textrm{QCD}$) effects.
The DY cross-section prediction accounts for varying
the strong coupling constant according to $\alpha_\textrm{s}(m_Z) = 0.118 \pm 0.002$,
a variation that corresponds to a 90\% CL uncertainty range~\cite{CT14}
that nevertheless has a small impact on the analysis.
Although the \ttbar\ cross-section uncertainty is only about 6\%~\cite{ATL-PHYS-PUB-2016-020}
and the corresponding impact on the total background is small, it is accounted for in
the statistical analysis due the characteristic \mt\ distribution shape for this background source.
An \mt-dependent uncertainty in the \ttbar\ shape is also included.
It corresponds to the remaining level of disagreement between the data and the simulation
after the correction described in Section~\ref{sec:simulation}. This uncertainty
is evaluated in a control region consisting of events with both an electron and a muon candidate,
which is a region dominated by \ttbar\ events.
The diboson cross-section uncertainty is neglected due to its small impact on the analysis.
However, the extrapolation uncertainty for the diboson background is included
in the statistical analysis as it grows to become significant at higher \mt\ values.
This uncertainty is estimated by varying the range of \mt\ values over which the fit is performed
and by changing the functional form. The extrapolation uncertainty for the top-quark background
is neglected due to its small impact.
 
The uncertainty in the integrated luminosity is 1.7\%~\cite{ATLAS-CONF-2019-021}.
 
Table~\ref{tab:syst} summarizes the systematic uncertainties for the total background and signal
in the electron and muon channels at \mt\ values near 2 and 6~\TeV.
The values in Table~\ref{tab:syst} correspond to the uncertainties that are incorporated as input to the statistical
analysis described in Section~\ref{sec:results}.
Large uncertainties in the background yields near \mt\ values of 6~\TeV\ are obtained but those
have little impact on the statistical analysis due to the small background expectation at such high
\mt\ values (e.g., the number of background events for $\mt > 5.1$~\TeV\ is 0.02 in the
electron channel and 0.11 in the muon channel).
 
\begin{table}
\caption{Systematic uncertainties in the expected number of events for the total background and for a \wp\ boson with a mass of 2~(6)~\TeV. The uncertainties are estimated with the binning shown in
Figure~\ref{fig:mt_panels} at \mt\ $=$ 2 (6)~\TeV\ for the background and in a three-bin window around \mt\ $=$ 2 (6)~\TeV\ for the signal.
Uncertainties that are not applicable are denoted ``{\textsc {n/a}}'', and ``negl.'' means that the uncertainty is not included in the statistical analysis because its impact on the result is negligible at any \mt\ value.
Small uncertainties that appear in the table (e.g. those listed as $< 0.5\%$) are not negligible at \mt\ values lower than 2~\TeV\ and are thus listed.
Sources of uncertainty not included in the table are neglected in the statistical analysis. \label{tab:syst}}
\begin{center}
\centering
\small
\begin{tabular}{l | r@{~~}l r@{~~}l | r@{~~}l r@{~~}l}
\toprule
\toprule
Source  &  \multicolumn{4}{c|}{Electron channel}  &  \multicolumn{4}{c}{Muon channel}  \\
& \multicolumn{2}{c}{Background}  & \multicolumn{2}{c|}{Signal}  & \multicolumn{2}{c}{Background} &  \multicolumn{2}{c}{Signal} \\
& \multicolumn{2}{c}{$\mt = 2~(6)$~\TeV} & \multicolumn{2}{c|}{$\mt = 2~(6)$~\TeV} & \multicolumn{2}{c}{$\mt = 2~(6)$~\TeV} & \multicolumn{2}{c}{$\mt = 2~(6)$~\TeV} \\
\midrule
Trigger &negl. & (negl.) & negl. & (negl.) & 1.1\% & (1.0\%) & 1.2\% & (1.2\%)\\
Lepton reconstruction & \multirow{2}{*}{4.1\%} & \multirow{2}{*}{(1.4\%)} & \multirow{2}{*}{4.3\%} & \multirow{2}{*}{(4.3\%)} & \multirow{2}{*}{8.9\%} & \multirow{2}{*}{(37\%)} & \multirow{2}{*}{6.6\%} & \multirow{2}{*}{(38\%)}\\
and identification & & & & & & & & \\
Lepton momentum &\multirow{2}{*}{3.9\%} & \multirow{2}{*}{(2.7\%)} & \multirow{2}{*}{2.7\%} & \multirow{2}{*}{(4.5\%)} &\multirow{2}{*}{12\%} & \multirow{2}{*}{(47\%)} & \multirow{2}{*}{13\%} & \multirow{2}{*}{(20\%)}\\
scale and resolution & & & & & & & & \\
\met{} resolution and scale &<0.5\% & (<0.5\%) &<0.5\% & (<0.5\%) &<0.5\% & (<0.5\%) &<0.5\% & (<0.5\%)\\
Jet energy resolution &<0.5\% & (<0.5\%) &<0.5\% & (<0.5\%) &<0.5\% & (0.6\%) &<0.5\% & (<0.5\%)\\
\midrule
Multijet background &4.4\% & (420\%) & {\textsc {n/a}} & ({\textsc {n/a}}) & 0.8\% & (1.5\%) & {\textsc {n/a}} & ({\textsc {n/a}})\\
Top-quark background &0.8\% & (1.9\%) & {\textsc {n/a}} & ({\textsc {n/a}}) &0.7\% & (<0.5\%) & {\textsc {n/a}} & ({\textsc {n/a}})\\
Diboson extrapolation &1.5\% & (47\%) & {\textsc {n/a}} & ({\textsc {n/a}}) &1.3\% & (9.7\%) & {\textsc {n/a}} & ({\textsc {n/a}})\\
PDF choice for DY &1.0\% & (10\%) & {\textsc {n/a}} & ({\textsc {n/a}}) &<0.5\% & (1.0\%) & {\textsc {n/a}} & ({\textsc {n/a}})\\
PDF variation for DY &8.1\% & (13\%) & {\textsc {n/a}} & ({\textsc {n/a}}) &7.4\% & (14\%) & {\textsc {n/a}} & ({\textsc {n/a}})\\
EW corrections for DY &4.2\% & (4.5\%) & {\textsc {n/a}} & ({\textsc {n/a}}) &3.7\% & (7.0\%) & {\textsc {n/a}} & ({\textsc {n/a}})\\
\midrule
Luminosity &1.6\% & (1.1\%) &1.7\% & (1.7\%) &1.7\% & (1.7\%) &1.7\% & (1.7\%)\\
\midrule
Total &12\% & (430\%) & 5.4\% & (6.4\%) &17\% & (62\%) & 15\% & (43\%)\\
\bottomrule
\bottomrule
\end{tabular}
\end{center}
\end{table}

\section{Results}
\label{sec:results}
 
The \mt\ distributions in the electron and muon channels (Figure~\ref{fig:mt_panels})
provide the input data to the statistical analysis.
This analysis proceeds as a multibin counting experiment with a likelihood accounting
for the Poisson probability to observe a number of events in data
given the expected number of background and signal events in each bin.
The uncertainties are taken into account via nuisance parameters implemented as
log-normal constraints on the expected event yields.
The parameter of interest is the cross section $\sigma(pp \to \wp \to \ell\nu)$.
The combined fits to the electron and muon channels are performed
taking correlations between the two channels into account.
 
The compatibility of the observed data with the background-only model is tested by computing a
frequentist $p$-value based on the profile likelihood ratio as the test statistic~\cite{Cowan:2010js}.
The $p$-value corresponds to the probability for the background to yield an excess equal to
or larger than that observed in data.
In the electron channel, the lowest $p$-value is obtained for $m(\wp) = 625$~\GeV\ with a local
significance of 2.8 standard deviations, corresponding to a global significance of 1.3 standard
deviations when taking the look-elsewhere effect into account.
In the muon channel, the lowest $p$-value is obtained for $m(\wp) = 200$~\GeV\ with local
and global significances of 2.1 and 0.4 standard deviations, respectively.
For the combination of the two channels, the lowest $p$-value occurs for $m(\wp) = 625$~\GeV\ with local
significance of 1.8 standard deviations, and the corresponding global significance is $-0.5$ standard
deviations (i.e.\ the fluctuation in the data is smaller than the median of the distribution obtained
with background-only pseudoexperiments).
In all cases, the interpretation is performed in the context of the SSM.
 
Given that no significant deviation from the background expectation is observed,
upper limits are set on $\sigma(pp \to \wp \to \ell\nu)$ following a Bayesian approach with a uniform
and positive prior for the cross section.
This choice of prior is the same as that used in previous searches~\cite{EXOT-2016-06,CMS-EXO-16-033}.
The marginalization of the posterior probability is performed using Markov Chain sampling
with the Bayesian Analysis Toolkit~\cite{2009CoPhC.180.2197C}.
Upper limits set at the 95\% CL in the context of the SSM are presented in Figure~\ref{fig:wprime_limits}
for the electron and muon channels individually as well as for their combination,
assuming universal \wp\ boson couplings to leptons.
The combined results are provided in terms of \wp\ boson decays into leptons of
a single generation.
The corresponding lower limits on the \wp\ boson mass are summarized in Table~\ref{tab:wprime_mass_limits}.
Weaker limits are obtained in the muon channel due to the lower signal acceptance times efficiency
and the worse momentum resolution at high \pt.
 
\begin{figure}[tb]
\centering
\includegraphics[width=0.59\textwidth]{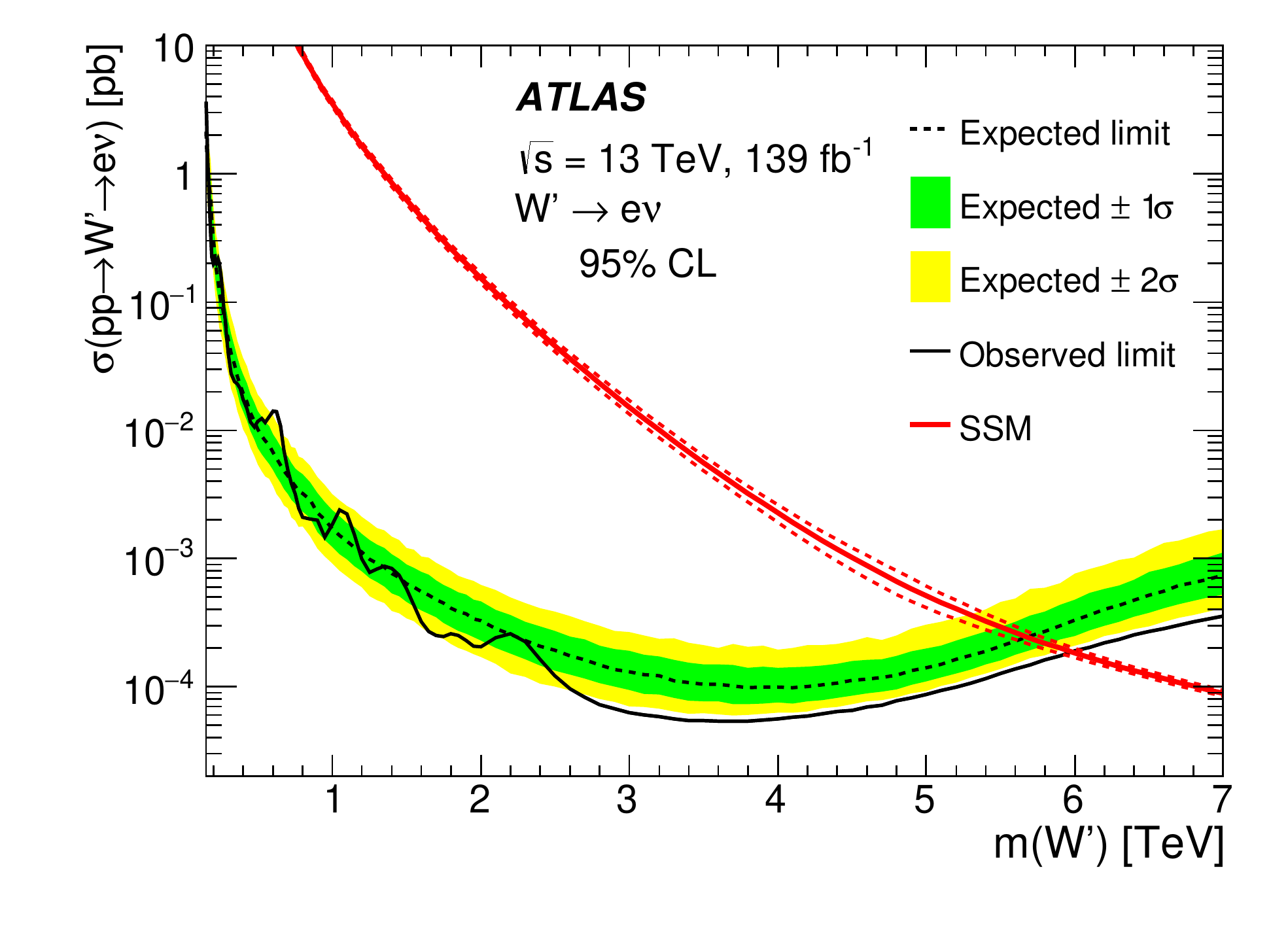}
\includegraphics[width=0.59\textwidth]{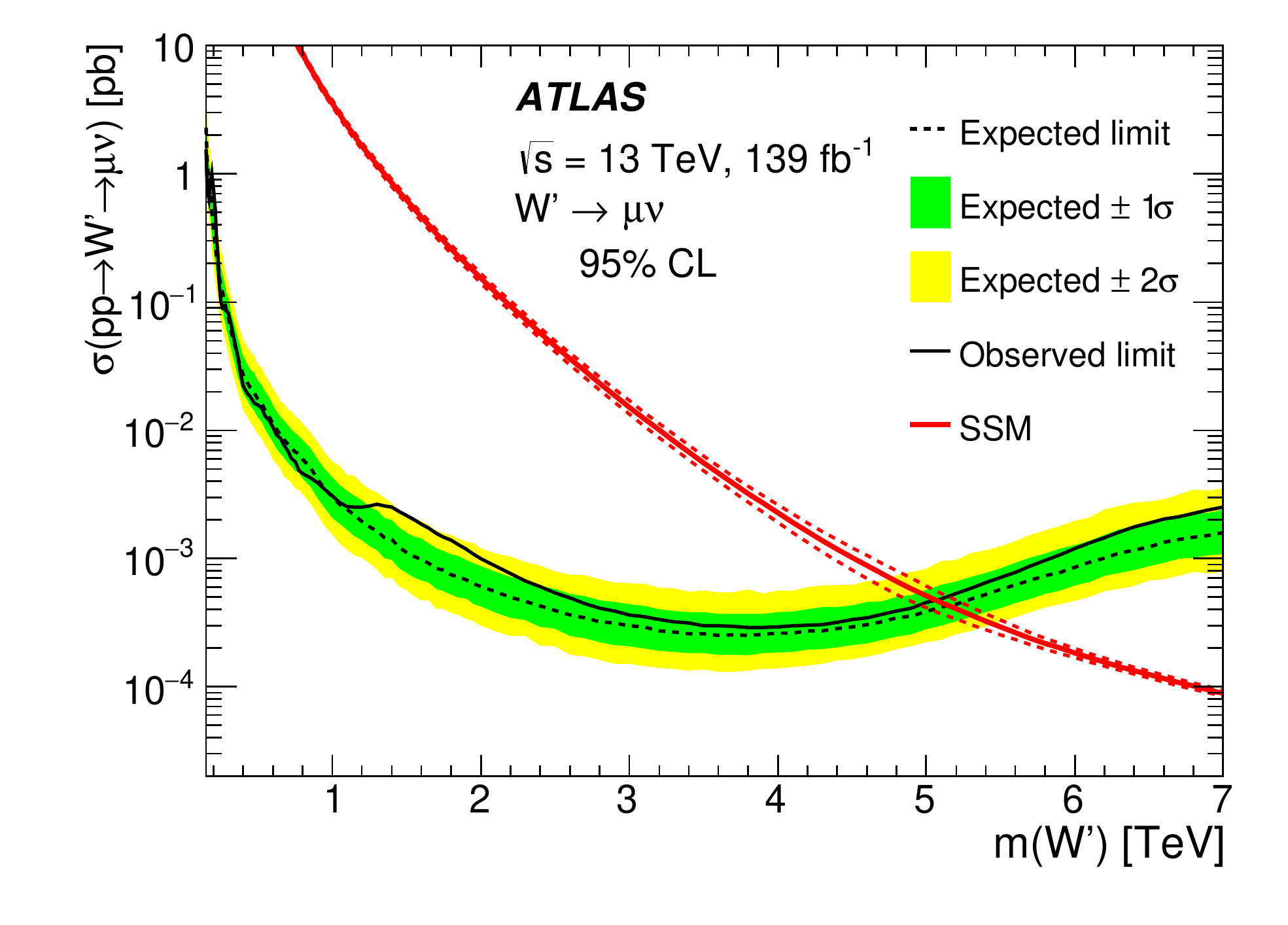}
\includegraphics[width=0.59\textwidth]{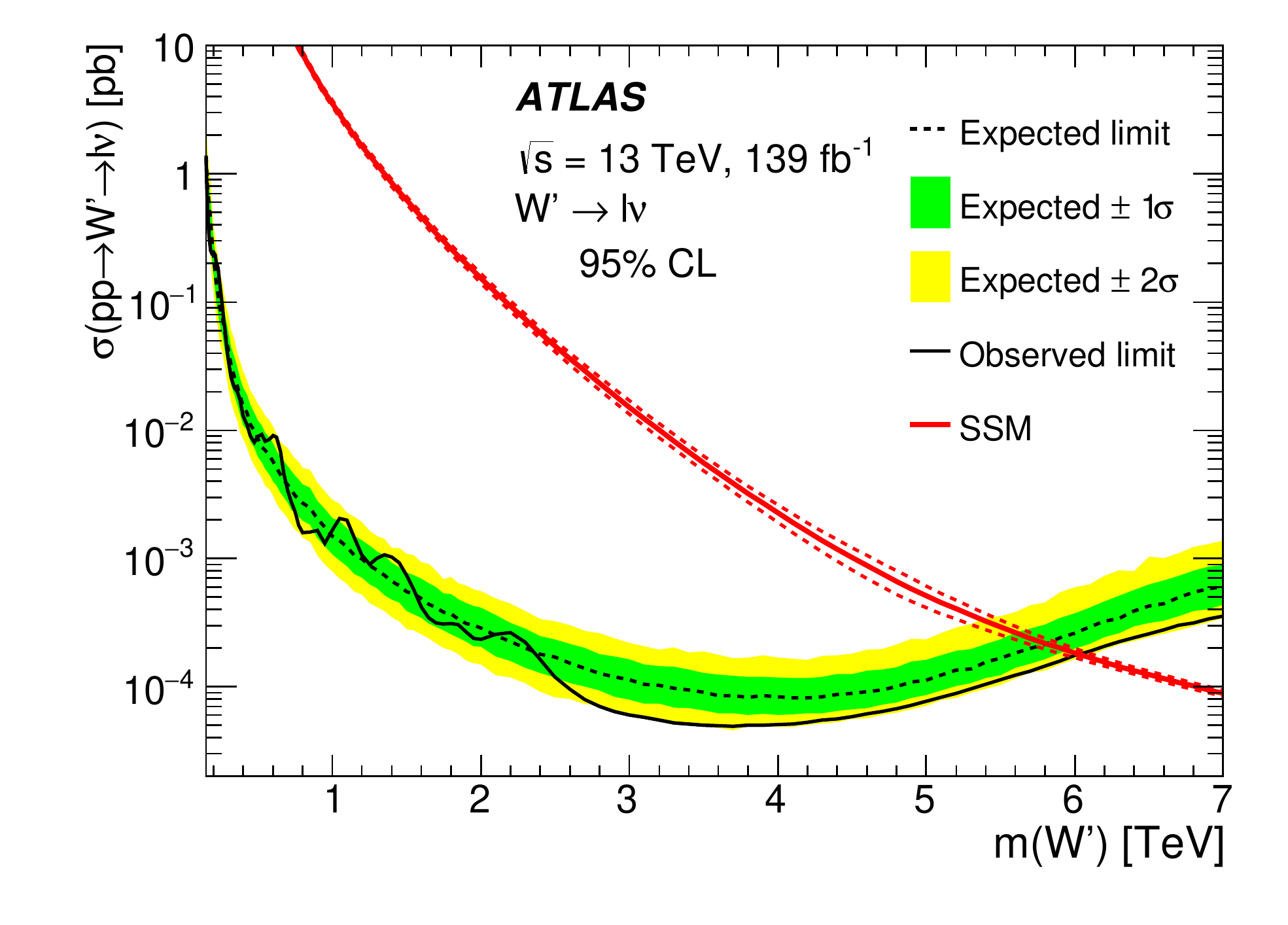}
\caption{Observed and expected upper limits at the 95\% CL on the $pp \to \wp \to \ell\nu$
cross section in the electron (top), muon (middle), and combined (bottom) channels
as a function of \wp\ mass in the Sequential Standard Model. The dashed lines surrounding the
SSM cross-section curve (solid line) correspond to the combination of PDF, $\alpha_\textrm{s}$,
renormalization, and factorization scale uncertainties (for illustration only).
}
\label{fig:wprime_limits}
\end{figure}
 
\begin{table}[tb]
\caption{Observed and expected 95\% CL lower limits on the \wp\ mass in the electron and muon
channels and their combination for the Sequential Standard Model.}
\centering
\begin{tabular}{ccc}
\hline
\hline
&  \multicolumn{2}{c}{$m(\wp)$ lower limit [\TeV]} \\
Decay     & Observed & Expected \\
\hline
\wpe  & 6.0 & 5.7 \\
\wpmu & 5.1 & 5.1 \\ \hline
\wpl  & 6.0 & 5.8 \\
\hline
\hline
\end{tabular}
\label{tab:wprime_mass_limits}
\end{table}
 
The lower panels of Figure~\ref{fig:mt_panels} show the ratio of the data to the background
prediction before (middle panel) and after (lower panel) marginalization of the nuisance parameters,
with the latter resulting from the combined fit to the electron and muon channels.
A difference in event yields is observed at low \mt\ values for both the electron and muon channels,
although it remains within the range of uncertainty before marginalization.
This difference decreases after marginalization, with the largest deviations from nominal values
occurring for the jet energy resolution and \met\ track soft term nuisance parameters.
The latter includes a significant model dependence found by comparing the predictions
from the \POWHEGBOX, \MGMCatNLO~\cite{madgraph}, and \SHERPA\ event generators, with the
first two interfaced with \PYTHIAV{8} for parton showering and hadronization.
 
The results displayed in Figure~\ref{fig:wprime_limits} are obtained with the full signal line shape
from the SSM with no interference between the \wp\ signal and the SM DY background.
If the signal line shape is restricted to the \wp\ peak region by the requirement
$m_{\ell\nu} > 0.85\times m(\wp)$, the interference effects in the low-mass tail of the
distributions are largely suppressed and the
observed (expected) mass limits become weaker by 270 (100)~\GeV\ in the electron channel
and 30 (90)~\GeV\ in the muon channel, relative to the mass limits
shown in Table~\ref{tab:wprime_mass_limits}.
The $m_{\ell\nu} > 0.85\times m(\wp)$ requirement is applied at the event generator level, considering
charged leptons before final-state radiation.
 
Limits are provided for the production of a generic resonance with a fixed $\Gamma / m$ value.
For these results, fiducial cross-section limits are obtained with a requirement that removes
the low-mass tail: $m_{\ell\nu} > 0.3\times m(\wp)$.
The region below $0.3\times m(\wp)$ coincides with the lower-\mt\ region where the background is large
and the sensitivity to signal contributions is reduced.
The observed 95\% CL upper limits on the fiducial cross section for $pp \to \wp \to \ell\nu$
with different choices of $\Gamma / m$ from 1\% to 15\% are shown in Figure~\ref{fig:generic_limits}.
Less stringent limits are obtained for larger resonance widths since a larger fraction of the
signal occurs in the low-\mt\ tail where the background is higher.
The cross section upper limits obtained in the fiducial region are lower than the ones obtained in
the full phase space, in particular at high $m(\wp)$ where the total cross section has a large
contribution from outside the fiducial region due to the low-\mt\ tail. The lower values of the cross
section limits do not indicate that the fiducial limits exclude a broader set of models,
as corresponding theoretical predictions are also lower in the fiducial than in the total phase space.
 
\begin{figure}[tb]
\centering
\includegraphics[width=0.59\textwidth]{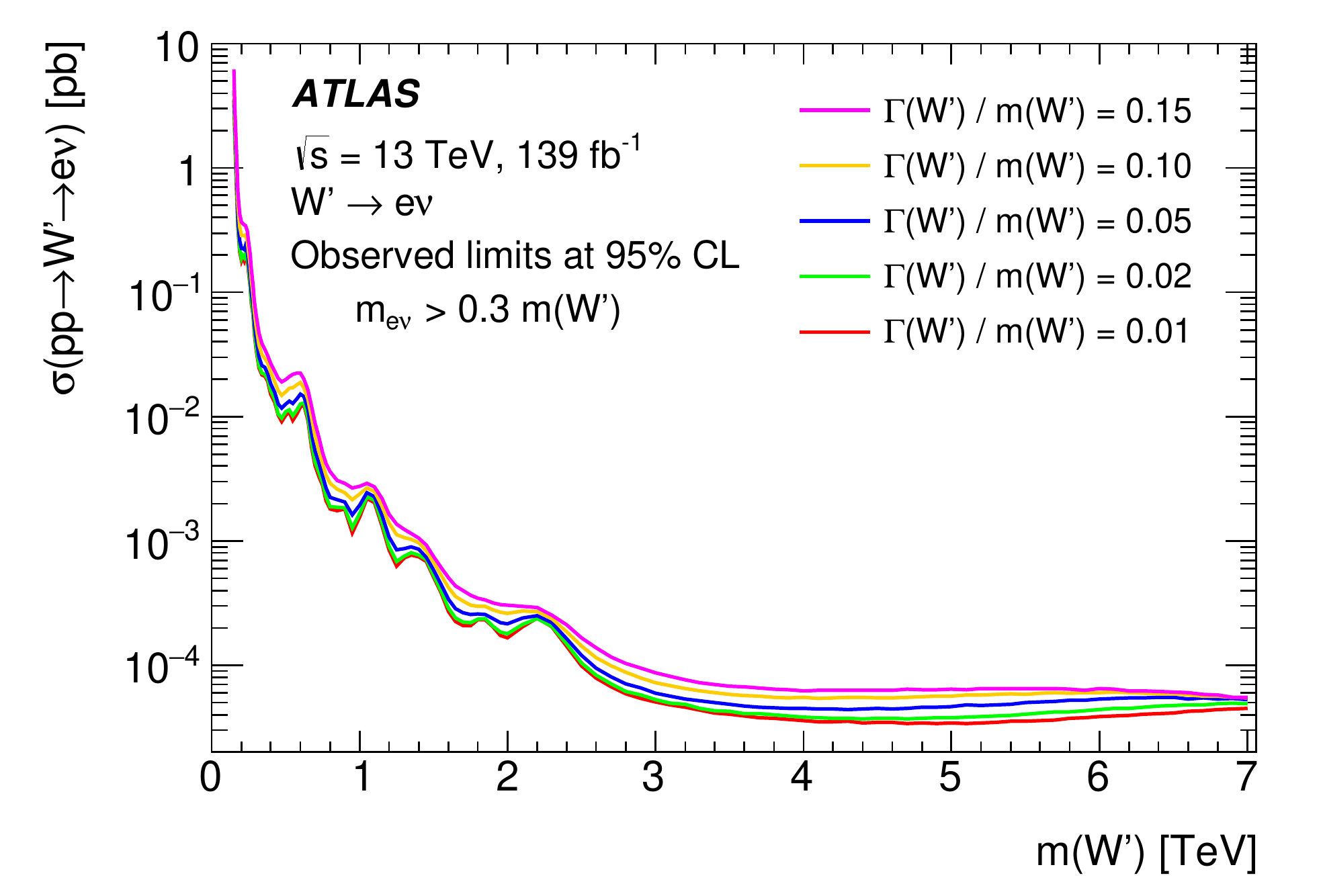}
\includegraphics[width=0.59\textwidth]{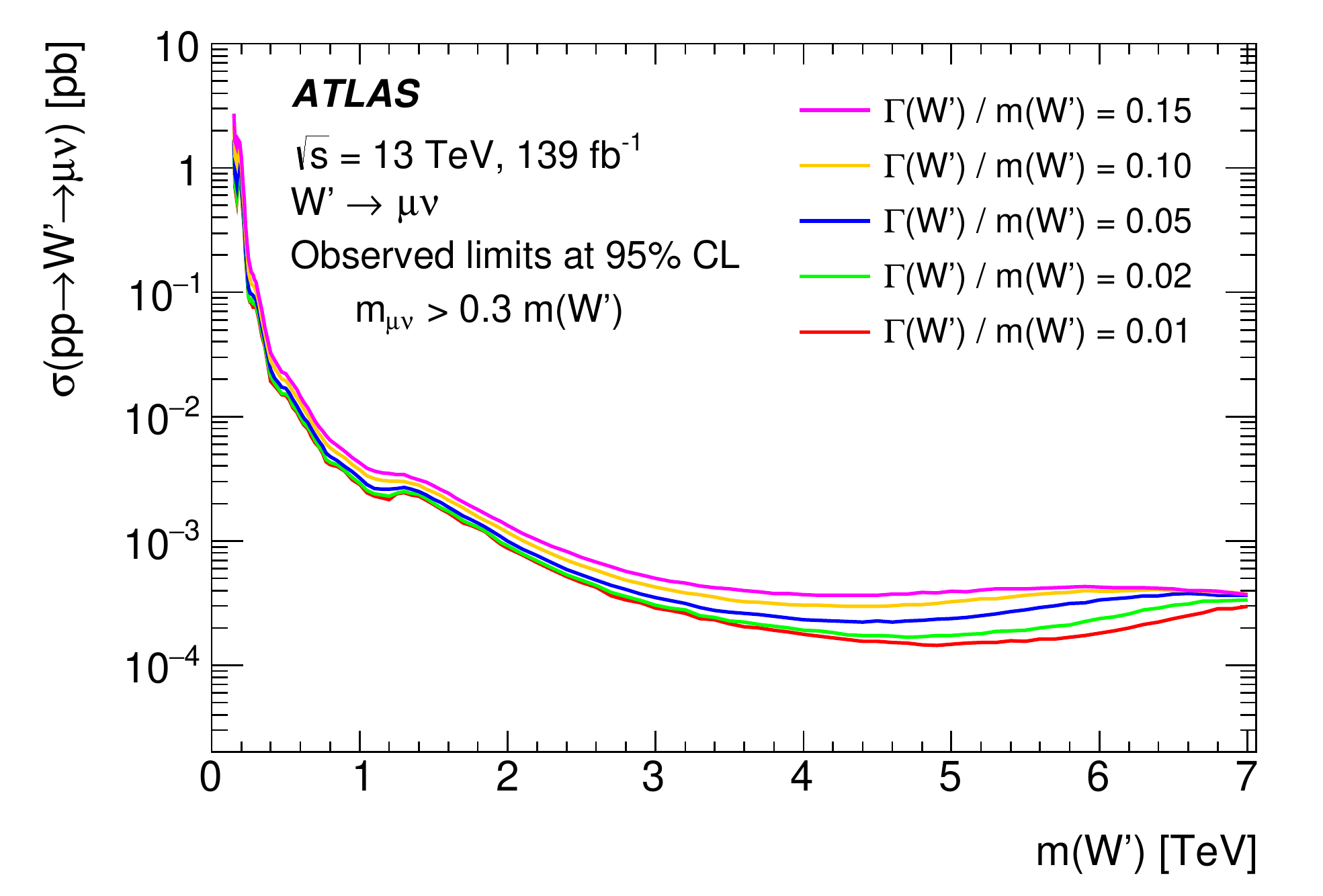}
\includegraphics[width=0.59\textwidth]{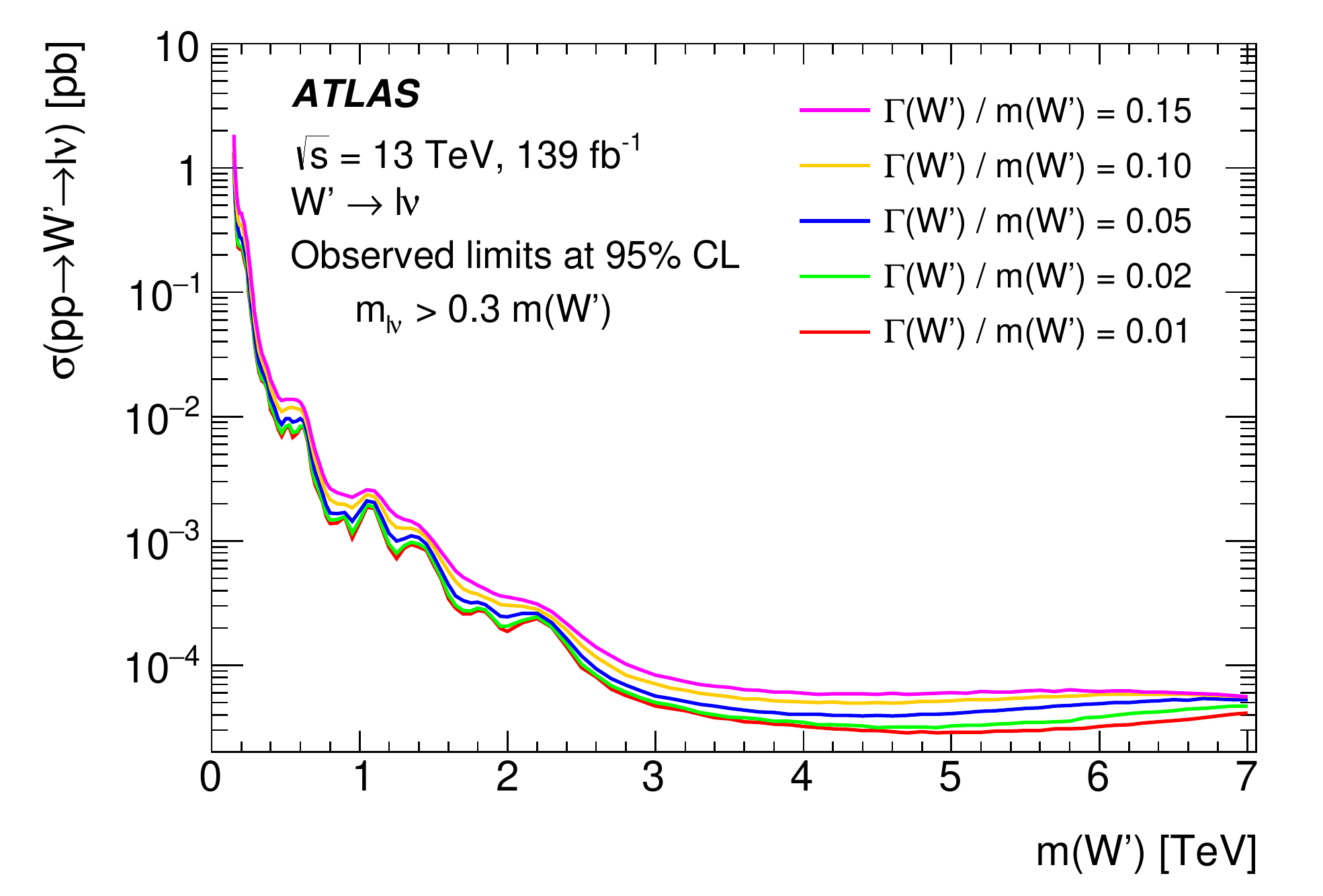}
\caption{Observed upper limits at the 95\% CL on the fiducial cross section for
$pp \to \wp \to \ell\nu$ in the electron (top), muon (middle), and combined (bottom) channels
as a function of \wp\ mass for a number of different choices of $\Gamma(\wp)/m(\wp)$ ranging between
1\% and 15\%.
}
\label{fig:generic_limits}
\end{figure}
 
To facilitate further interpretations of the results, model-independent upper limits are also provided for the number
of signal events $N_\mathrm{sig}$ in single-bin signal regions obtained by varying the minimum \mt\ value \mtmin\ in
the range between 130 (110)~\GeV\ and 5127 (5127)~\GeV\ in the electron (muon) channel.
These limits are translated into limits on the visible cross section $\sigma_\mathrm{vis}$ computed
as $N_\mathrm{sig}/\cal{L}$, where $\cal{L}$ is the integrated luminosity.
The visible cross section corresponds to the product of cross section times acceptance times efficiency
and the observed 95\% CL upper limits vary from 4.6 (15)~pb at $\mtmin = 130$ (110)~\GeV\ to 22 (22)~ab at high \mtmin\
in the electron (muon) channel as shown in Figure~\ref{fig:modelIndepLimits}.
Further details about these model-independent limits are available in the Appendix.
 
\begin{figure}[tb]
\centering
\includegraphics[width=0.59\linewidth]{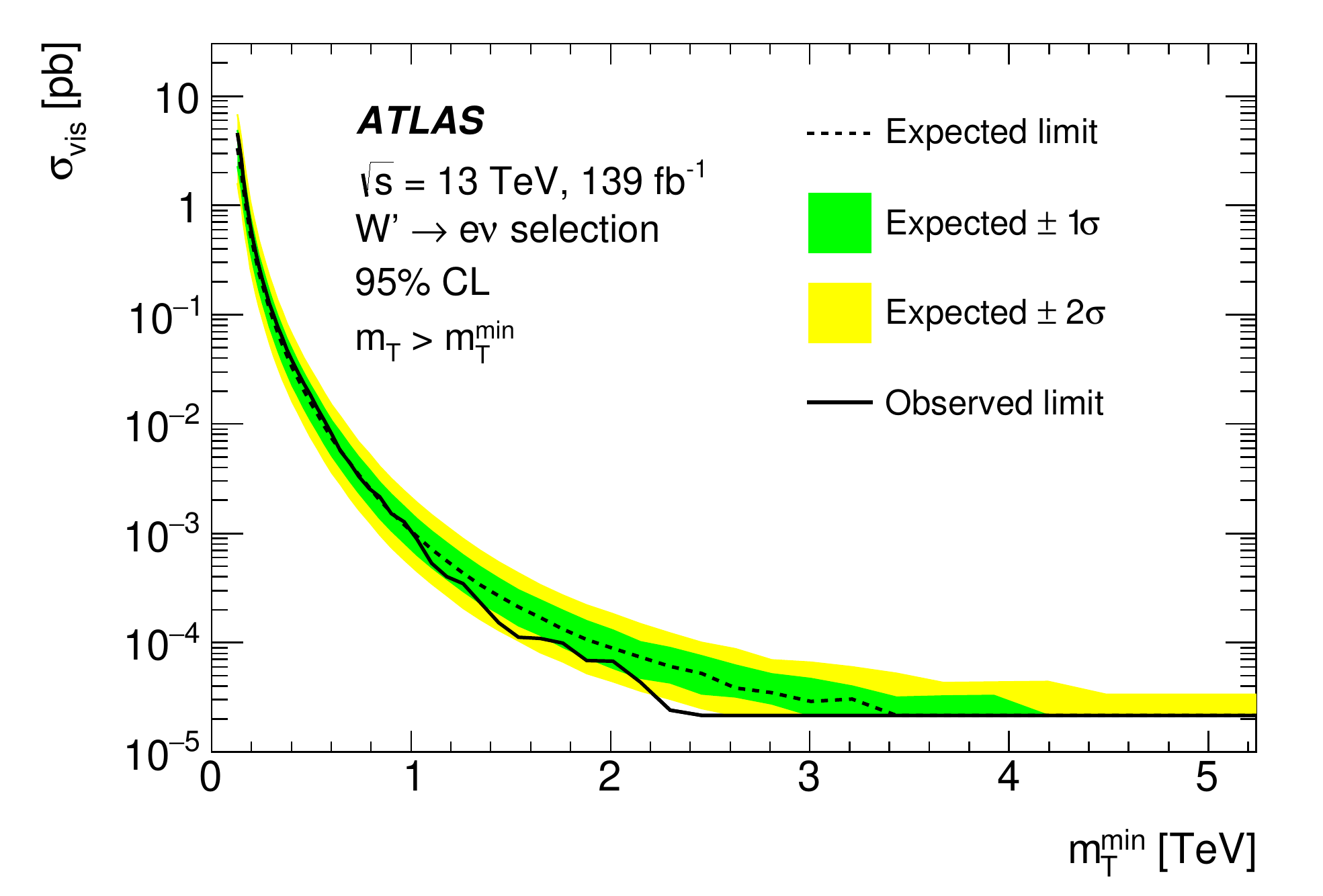}
\includegraphics[width=0.59\linewidth]{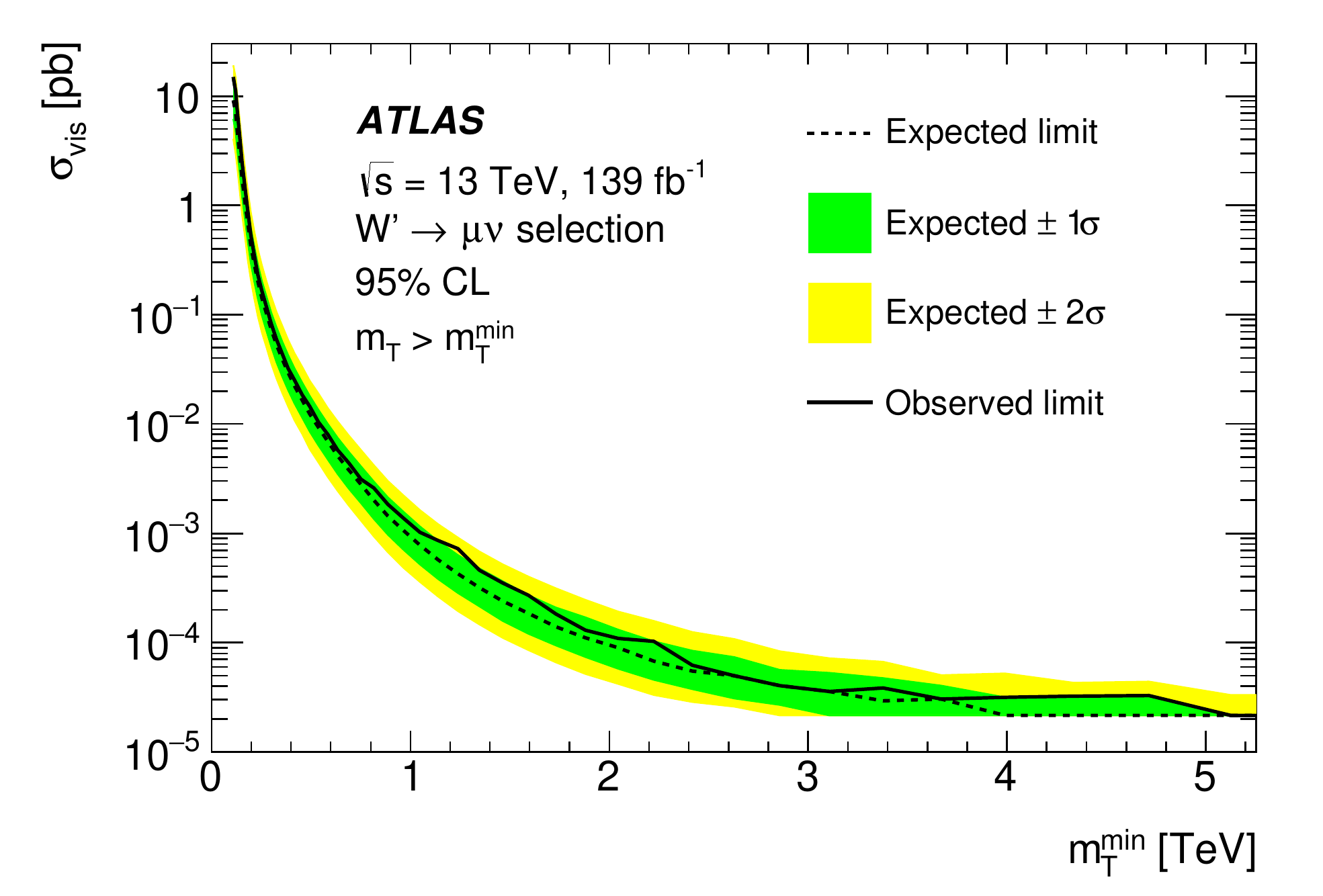}
\caption{Observed and expected model-independent upper limits at the 95\% CL on the visible cross section
in the electron (top) and muon (bottom) channels as a function of the \mt\ threshold \mtmin.
The limits are obtained at discrete \mtmin\ values and are connected by a straight line
for illustration purposes.}
\label{fig:modelIndepLimits}
\end{figure}
 
\FloatBarrier
 
\section{Conclusion}
 
A search for a heavy resonance decaying into a charged lepton and a neutrino
is carried out in events with an isolated electron or muon and missing transverse momentum.
The data sample corresponds to \totlumi~\ifb\ of $pp$ collisions at $\rts = 13$~\TeV\
collected in 2015--2018 with the ATLAS detector at the LHC.
Events are selected with single-electron and single-muon triggers, and the transverse
mass computed from the lepton \pt\ and the missing transverse momentum is used as the
discriminating variable between signal and background contributions.
The latter is dominated by Drell--Yan production of $W$ bosons.
Monte Carlo simulation is used to estimate the normalization and shape of the \mt\
distributions for signal and background events, except for the multijet background,
which is derived from the data.
 
The observed \mt\ distributions are found to be consistent with the background expectations,
and upper limits are set on the cross section for $pp \to \wp \to \ell\nu$,
where the charged lepton is either an electron or a muon. Limits are also provided
for the combination of the electron and muon channels.
Lower limits of 6.0~\TeV\ and 5.1~\TeV\ on the \wp\ boson mass are set at 95\% CL in the
electron and muon channels, respectively, in the context of the Sequential Standard Model.
Fiducial cross-section limits are set on the production of resonances with different $\Gamma/m$ values
ranging from 1\% to 15\%. To allow for further interpretations of the results,
a set of model-independent upper limits are presented for the number of signal
events and for the visible cross section above a given transverse mass threshold.
These vary from 4.6 (15)~pb at $\mtmin = 130$ (110)~\GeV\ to 22 (22)~ab at high \mtmin\
in the electron (muon) channel.

\section*{Acknowledgments}
 
 
We thank CERN for the very successful operation of the LHC, as well as the
support staff from our institutions without whom ATLAS could not be
operated efficiently.
 
We acknowledge the support of ANPCyT, Argentina; YerPhI, Armenia; ARC, Australia; BMWFW and FWF, Austria; ANAS, Azerbaijan; SSTC, Belarus; CNPq and FAPESP, Brazil; NSERC, NRC and CFI, Canada; CERN; CONICYT, Chile; CAS, MOST and NSFC, China; COLCIENCIAS, Colombia; MSMT CR, MPO CR and VSC CR, Czech Republic; DNRF and DNSRC, Denmark; IN2P3-CNRS, CEA-DRF/IRFU, France; SRNSFG, Georgia; BMBF, HGF, and MPG, Germany; GSRT, Greece; RGC, Hong Kong SAR, China; ISF and Benoziyo Center, Israel; INFN, Italy; MEXT and JSPS, Japan; CNRST, Morocco; NWO, Netherlands; RCN, Norway; MNiSW and NCN, Poland; FCT, Portugal; MNE/IFA, Romania; MES of Russia and NRC KI, Russian Federation; JINR; MESTD, Serbia; MSSR, Slovakia; ARRS and MIZ\v{S}, Slovenia; DST/NRF, South Africa; MINECO, Spain; SRC and Wallenberg Foundation, Sweden; SERI, SNSF and Cantons of Bern and Geneva, Switzerland; MOST, Taiwan; TAEK, Turkey; STFC, United Kingdom; DOE and NSF, United States of America. In addition, individual groups and members have received support from BCKDF, CANARIE, CRC and Compute Canada, Canada; COST, ERC, ERDF, Horizon 2020, and Marie Sk{l}odowska-Curie Actions, European Union; Investissements d' Avenir Labex and Idex, ANR, France; DFG and AvH Foundation, Germany; Herakleitos, Thales and Aristeia programmes co-financed by EU-ESF and the Greek NSRF, Greece; BSF-NSF and GIF, Israel; CERCA Programme Generalitat de Catalunya, Spain; The Royal Society and Leverhulme Trust, United Kingdom.
 
The crucial computing support from all WLCG partners is acknowledged gratefully, in particular from CERN, the ATLAS Tier-1 facilities at TRIUMF (Canada), NDGF (Denmark, Norway, Sweden), CC-IN2P3 (France), KIT/GridKA (Germany), INFN-CNAF (Italy), NL-T1 (Netherlands), PIC (Spain), ASGC (Taiwan), RAL (UK) and BNL (USA), the Tier-2 facilities worldwide and large non-WLCG resource providers. Major contributors of computing resources are listed in Ref.~\cite{ATL-GEN-PUB-2016-002}.
 

\clearpage
\appendix
\part*{Appendix}
 
Model-independent upper limits are derived by applying the full event selection in a set of single-bin
signal regions defined by the minimum \mt\ value \mtmin\ in the range between 130 (110)~\GeV\ and 5127 (5127)~\GeV,
in the electron (muon) channel.
These minimum values correspond to the bin boundaries of the \mt\ distributions shown in Figure~\ref{fig:mt_panels}.
The single-bin signal regions are defined in Tables~\ref{tab:modelIndepLimits_el} and \ref{tab:modelIndepLimits_mu}.
These tables also show the numbers of events observed in data and the expected numbers of background events.
 
\begin{table}[tbp]
\centering
\caption{Observed and expected electron-channel model-independent limits at 95\% CL on the number of signal events $N_\mathrm{sig}$ and corresponding visible cross section
$\sigma_\mathrm{vis}$ after full event selection for different \mt\ thresholds $m_\mathrm{T}^\mathrm{min}$. Also shown are the ingredients
to the limit calculation, namely the number of observed events, the expected number of background events $b$, and the corresponding uncertainty $\Delta_b$.}
\scriptsize
\begin{tabular}{|c|c|c|c|c|c|c|c|}
\hline
$m_\mathrm{T}^\mathrm{min}$ [\GeV] & $N_\mathrm{obs}$ & $b$ & $\Delta_b$ & \multicolumn{4}{c|}{Upper limit at 95\% CL} \\ \cline{5-8}
&  &  &  & $N_\mathrm{sig}^\mathrm{obs}$ & $N_\mathrm{sig}^\mathrm{exp}$ &
$\sigma_\mathrm{vis}^\mathrm{obs}$ [pb] & $\sigma_\mathrm{vis}^\mathrm{exp}$ [pb] \\ \hline
$ 130 $ &  $ 3\,582\,164 $ &  $ 3\,360\,000 $ &  $ 250\,000 $ &  $ 6.4\times10^5 $ &  $ 4.6\times10^5 $ &  $ 4.6 $ &  $ 3.3 $ \\ \hline
$ 139 $ &  $ 3\,018\,934 $ &  $ 2\,850\,000 $ &  $ 200\,000 $ &  $ 5.1\times10^5 $ &  $ 3.8\times10^5 $ &  $ 3.7 $ &  $ 2.7 $ \\ \hline
$ 149 $ &  $ 2\,345\,269 $ &  $ 2\,240\,000 $ &  $ 150\,000 $ &  $ 3.6\times10^5 $ &  $ 2.8\times10^5 $ &  $ 2.6 $ &  $ 2.0 $ \\ \hline
$ 159 $ &  $ 1\,784\,938 $ &  $ 1\,720\,000 $ &  $ 110\,000 $ &  $ 2.5\times10^5 $ &  $ 2.0\times10^5 $ &  $ 1.8 $ &  $ 1.4 $ \\ \hline
$ 170 $ &  $ 1\,352\,988 $ &  $ 1\,310\,000 $ &  $ 80\,000 $ &  $ 1.7\times10^5 $ &  $ 1.4\times10^5 $ &  $ 1.3 $ &  $ 1.0 $ \\ \hline
$ 182 $ &  $ 1\,028\,353 $ &  $ 1\,000\,000 $ &  $ 60\,000 $ &  $ 1.2\times10^5 $ &  $ 1.1\times10^5 $ &  $ 0.90 $ &  $ 0.76 $ \\ \hline
$ 194 $ &  $ 784\,509 $ &  $ 770\,000 $ &  $ 40\,000 $ &  $ 9.1\times10^4 $ &  $ 7.7\times10^4 $ &  $ 0.66 $ &  $ 0.55 $ \\ \hline
$ 208 $ &  $ 599\,989 $ &  $ 588\,000 $ &  $ 31\,000 $ &  $ 6.7\times10^4 $ &  $ 5.8\times10^4 $ &  $ 0.48 $ &  $ 0.42 $ \\ \hline
$ 222 $ &  $ 459\,843 $ &  $ 451\,000 $ &  $ 23\,000 $ &  $ 5.0\times10^4 $ &  $ 4.4\times10^4 $ &  $ 0.36 $ &  $ 0.31 $ \\ \hline
$ 237 $ &  $ 352\,825 $ &  $ 347\,000 $ &  $ 18\,000 $ &  $ 3.8\times10^4 $ &  $ 3.4\times10^4 $ &  $ 0.27 $ &  $ 0.24 $ \\ \hline
$ 254 $ &  $ 270\,299 $ &  $ 267\,000 $ &  $ 14\,000 $ &  $ 2.9\times10^4 $ &  $ 2.6\times10^4 $ &  $ 0.21 $ &  $ 0.19 $ \\ \hline
$ 271 $ &  $ 207\,728 $ &  $ 204\,000 $ &  $ 11\,000 $ &  $ 2.3\times10^4 $ &  $ 2.0\times10^4 $ &  $ 0.16 $ &  $ 0.15 $ \\ \hline
$ 290 $ &  $ 159\,319 $ &  $ 157\,000 $ &  $ 8000 $ &  $ 1.7\times10^4 $ &  $ 1.6\times10^4 $ &  $ 0.13 $ &  $ 0.11 $ \\ \hline
$ 310 $ &  $ 122\,150 $ &  $ 120\,000 $ &  $ 6000 $ &  $ 1.4\times10^4 $ &  $ 1.2\times10^4 $ &  $ 0.10 $ &  $ 0.088 $ \\ \hline
$ 331 $ &  $ 93\,335 $ &  $ 92\,000 $ &  $ 5000 $ &  $ 1.1\times10^4 $ &  $ 9.5\times10^3 $ &  $ 0.078 $ &  $ 0.069 $ \\ \hline
$ 354 $ &  $ 71\,416 $ &  $ 70\,000 $ &  $ 4000 $ &  $ 8.6\times10^3 $ &  $ 7.4\times10^3 $ &  $ 0.062 $ &  $ 0.053 $ \\ \hline
$ 379 $ &  $ 54\,642 $ &  $ 53\,500 $ &  $ 3100 $ &  $ 6.6\times10^3 $ &  $ 5.8\times10^3 $ &  $ 0.048 $ &  $ 0.042 $ \\ \hline
$ 405 $ &  $ 41\,745 $ &  $ 40\,800 $ &  $ 2400 $ &  $ 5.3\times10^3 $ &  $ 4.5\times10^3 $ &  $ 0.038 $ &  $ 0.033 $ \\ \hline
$ 433 $ &  $ 31\,792 $ &  $ 31\,100 $ &  $ 1900 $ &  $ 4.1\times10^3 $ &  $ 3.6\times10^3 $ &  $ 0.030 $ &  $ 0.026 $ \\ \hline
$ 463 $ &  $ 24\,257 $ &  $ 23\,600 $ &  $ 1500 $ &  $ 3.3\times10^3 $ &  $ 2.8\times10^3 $ &  $ 0.023 $ &  $ 0.020 $ \\ \hline
$ 495 $ &  $ 18\,484 $ &  $ 18\,000 $ &  $ 1200 $ &  $ 2.6\times10^3 $ &  $ 2.2\times10^3 $ &  $ 0.019 $ &  $ 0.016 $ \\ \hline
$ 529 $ &  $ 13\,937 $ &  $ 13\,600 $ &  $ 900 $ &  $ 1.9\times10^3 $ &  $ 1.7\times10^3 $ &  $ 0.014 $ &  $ 0.012 $ \\ \hline
$ 565 $ &  $ 10\,548 $ &  $ 10\,300 $ &  $ 700 $ &  $ 1.5\times10^3 $ &  $ 1.3\times10^3 $ &  $ 0.011 $ &  $ 0.0096 $ \\ \hline
$ 604 $ &  $ 7938 $ &  $ 7800 $ &  $ 500 $ &  $ 1.1\times10^3 $ &  $ 1.0\times10^3 $ &  $ 0.0080 $ &  $ 0.0074 $ \\ \hline
$ 646 $ &  $ 5926 $ &  $ 5900 $ &  $ 400 $ &  $ 7.8\times10^2 $ &  $ 8.0\times10^2 $ &  $ 0.0056 $ &  $ 0.0057 $ \\ \hline
$ 691 $ &  $ 4469 $ &  $ 4470 $ &  $ 330 $ &  $ 6.2\times10^2 $ &  $ 6.2\times10^2 $ &  $ 0.0044 $ &  $ 0.0044 $ \\ \hline
$ 739 $ &  $ 3342 $ &  $ 3360 $ &  $ 250 $ &  $ 4.6\times10^2 $ &  $ 4.8\times10^2 $ &  $ 0.0033 $ &  $ 0.0034 $ \\ \hline
$ 790 $ &  $ 2499 $ &  $ 2510 $ &  $ 190 $ &  $ 3.6\times10^2 $ &  $ 3.7\times10^2 $ &  $ 0.0026 $ &  $ 0.0026 $ \\ \hline
$ 844 $ &  $ 1876 $ &  $ 1850 $ &  $ 140 $ &  $ 3.0\times10^2 $ &  $ 2.8\times10^2 $ &  $ 0.0022 $ &  $ 0.0020 $ \\ \hline
$ 902 $ &  $ 1358 $ &  $ 1370 $ &  $ 110 $ &  $ 2.1\times10^2 $ &  $ 2.2\times10^2 $ &  $ 0.0015 $ &  $ 0.0016 $ \\ \hline
$ 965 $ &  $ 1021 $ &  $ 1010 $ &  $ 80 $ &  $ 1.8\times10^2 $ &  $ 1.7\times10^2 $ &  $ 0.0013 $ &  $ 0.0012 $ \\ \hline
$ 1031 $ &  $ 727 $ &  $ 740 $ &  $ 60 $ &  $ 1.2\times10^2 $ &  $ 1.3\times10^2 $ &  $ 0.00088 $ &  $ 0.00093 $ \\ \hline
$ 1103 $ &  $ 495 $ &  $ 540 $ &  $ 50 $ &  $ 74 $ &  $ 1.0\times10^2 $ &  $ 0.00053 $ &  $ 0.00072 $ \\ \hline
$ 1179 $ &  $ 354 $ &  $ 390 $ &  $ 40 $ &  $ 56 $ &  $ 78 $ &  $ 0.00040 $ &  $ 0.00056 $ \\ \hline
$ 1260 $ &  $ 260 $ &  $ 278 $ &  $ 27 $ &  $ 48 $ &  $ 60 $ &  $ 0.00035 $ &  $ 0.00043 $ \\ \hline
$ 1347 $ &  $ 175 $ &  $ 198 $ &  $ 20 $ &  $ 33 $ &  $ 47 $ &  $ 0.00024 $ &  $ 0.00034 $ \\ \hline
$ 1441 $ &  $ 113 $ &  $ 140 $ &  $ 15 $ &  $ 21 $ &  $ 37 $ &  $ 0.00015 $ &  $ 0.00027 $ \\ \hline
$ 1540 $ &  $ 74 $ &  $ 98 $ &  $ 11 $ &  $ 16 $ &  $ 29 $ &  $ 0.00011 $ &  $ 0.00021 $ \\ \hline
$ 1647 $ &  $ 55 $ &  $ 68 $ &  $ 8 $ &  $ 15 $ &  $ 24 $ &  $ 0.00011 $ &  $ 0.00017 $ \\ \hline
$ 1760 $ &  $ 39 $ &  $ 46 $ &  $ 6 $ &  $ 14 $ &  $ 19 $ &  $ 9.9\times10^{-5} $ &  $ 0.00013 $ \\ \hline
$ 1882 $ &  $ 23 $ &  $ 31 $ &  $ 5 $ &  $ 9.6 $ &  $ 15 $ &  $ 6.9\times10^{-5} $ &  $ 0.00011 $ \\ \hline
$ 2012 $ &  $ 17 $ &  $ 20.9 $ &  $ 3.4 $ &  $ 9.4 $ &  $ 12 $ &  $ 6.8\times10^{-5} $ &  $ 8.9\times10^{-5} $ \\ \hline
$ 2151 $ &  $ 8 $ &  $ 13.7 $ &  $ 2.5 $ &  $ 6.0 $ &  $ 10 $ &  $ 4.3\times10^{-5} $ &  $ 7.4\times10^{-5} $ \\ \hline
$ 2300 $ &  $ 1 $ &  $ 8.9 $ &  $ 1.8 $ &  $ 3.4 $ &  $ 8.4 $ &  $ 2.4\times10^{-5} $ &  $ 6.1\times10^{-5} $ \\ \hline
$ 2458 $ &  $ 0 $ &  $ 5.7 $ &  $ 1.4 $ &  $ 3.0 $ &  $ 7.3 $ &  $ 2.2\times10^{-5} $ &  $ 5.2\times10^{-5} $ \\ \hline
$ 2628 $ &  $ 0 $ &  $ 3.6 $ &  $ 1.0 $ &  $ 3.0 $ &  $ 5.3 $ &  $ 2.2\times10^{-5} $ &  $ 3.8\times10^{-5} $ \\ \hline
$ 2810 $ &  $ 0 $ &  $ 2.2 $ &  $ 0.8 $ &  $ 3.0 $ &  $ 4.9 $ &  $ 2.2\times10^{-5} $ &  $ 3.5\times10^{-5} $ \\ \hline
$ 3004 $ &  $ 0 $ &  $ 1.3 $ &  $ 0.6 $ &  $ 3.0 $ &  $ 4.1 $ &  $ 2.2\times10^{-5} $ &  $ 2.9\times10^{-5} $ \\ \hline
$ 3212 $ &  $ 0 $ &  $ 0.8 $ &  $ 0.5 $ &  $ 3.0 $ &  $ 4.2 $ &  $ 2.2\times10^{-5} $ &  $ 3.1\times10^{-5} $ \\ \hline
$ 3434 $ &  $ 0 $ &  $ 0.5 $ &  $ 0.4 $ &  $ 3.0 $ &  $ 3.0 $ &  $ 2.2\times10^{-5} $ &  $ 2.2\times10^{-5} $ \\ \hline
$ 3671 $ &  $ 0 $ &  $ 0.28 $ &  $ 0.28 $ &  $ 3.0 $ &  $ 3.0 $ &  $ 2.2\times10^{-5} $ &  $ 2.2\times10^{-5} $ \\ \hline
$ 3924 $ &  $ 0 $ &  $ 0.16 $ &  $ 0.22 $ &  $ 3.0 $ &  $ 3.0 $ &  $ 2.2\times10^{-5} $ &  $ 2.2\times10^{-5} $ \\ \hline
$ 4196 $ &  $ 0 $ &  $ 0.09 $ &  $ 0.17 $ &  $ 3.0 $ &  $ 3.0 $ &  $ 2.2\times10^{-5} $ &  $ 2.2\times10^{-5} $ \\ \hline
$ 4485 $ &  $ 0 $ &  $ 0.05 $ &  $ 0.13 $ &  $ 3.0 $ &  $ 3.0 $ &  $ 2.2\times10^{-5} $ &  $ 2.2\times10^{-5} $ \\ \hline
$ 4795 $ &  $ 0 $ &  $ 0.03 $ &  $ 0.10 $ &  $ 3.0 $ &  $ 3.0 $ &  $ 2.2\times10^{-5} $ &  $ 2.2\times10^{-5} $ \\ \hline
$ 5127 $ &  $ 0 $ &  $ 0.02 $ &  $ 0.08 $ &  $ 3.0 $ &  $ 3.0 $ &  $ 2.2\times10^{-5} $ &  $ 2.2\times10^{-5} $ \\ \hline
\end{tabular}
\label{tab:modelIndepLimits_el}
\end{table}
 
\begin{table}[tbp]
\centering
\caption{Observed and expected muon-channel model-independent limits at 95\% CL on the number of signal events $N_\mathrm{sig}$ and corresponding visible cross section
$\sigma_\mathrm{vis}$ after full event selection for different \mt\ thresholds $m_\mathrm{T}^\mathrm{min}$. Also shown are the ingredients
to the limit calculation, namely the number of observed events, the expected number of background events $b$, and the corresponding uncertainty $\Delta_b$.}
\scriptsize
\begin{tabular}{|c|c|c|c|c|c|c|c|}
\hline
$m_\mathrm{T}^\mathrm{min}$ [\GeV] & $N_\mathrm{obs}$ & $b$ & $\Delta_b$ & \multicolumn{4}{c|}{Upper limit at 95\% CL} \\ \cline{5-8}
&  &  &  & $N_\mathrm{sig}^\mathrm{obs}$ & $N_\mathrm{sig}^\mathrm{exp}$ &
$\sigma_\mathrm{vis}^\mathrm{obs}$ [pb] & $\sigma_\mathrm{vis}^\mathrm{exp}$ [pb] \\ \hline
$ 110 $ &  $ 8\,783\,359 $ &  $ 7\,800\,000 $ &  $ 700\,000 $ &  $ 2.1\times10^6 $ &  $ 1.3\times10^6 $ &  $ 15 $ &  $ 9.1 $ \\ \hline
$ 120 $ &  $ 6\,589\,361 $ &  $ 5\,900\,000 $ &  $ 500\,000 $ &  $ 1.5\times10^6 $ &  $ 9.8\times10^5 $ &  $ 11 $ &  $ 7.0 $ \\ \hline
$ 130 $ &  $ 4\,353\,441 $ &  $ 3\,900\,000 $ &  $ 400\,000 $ &  $ 9.9\times10^5 $ &  $ 6.5\times10^5 $ &  $ 7.1 $ &  $ 4.7 $ \\ \hline
$ 141 $ &  $ 2\,820\,607 $ &  $ 2\,590\,000 $ &  $ 220\,000 $ &  $ 5.9\times10^5 $ &  $ 4.1\times10^5 $ &  $ 4.3 $ &  $ 2.9 $ \\ \hline
$ 154 $ &  $ 1\,840\,357 $ &  $ 1\,720\,000 $ &  $ 140\,000 $ &  $ 3.5\times10^5 $ &  $ 2.5\times10^5 $ &  $ 2.5 $ &  $ 1.8 $ \\ \hline
$ 167 $ &  $ 1\,227\,452 $ &  $ 1\,160\,000 $ &  $ 80\,000 $ &  $ 2.0\times10^5 $ &  $ 1.5\times10^5 $ &  $ 1.5 $ &  $ 1.1 $ \\ \hline
$ 182 $ &  $ 837\,724 $ &  $ 800\,000 $ &  $ 50\,000 $ &  $ 1.2\times10^5 $ &  $ 9.3\times10^4 $ &  $ 0.88 $ &  $ 0.67 $ \\ \hline
$ 197 $ &  $ 581\,304 $ &  $ 562\,000 $ &  $ 32\,000 $ &  $ 7.5\times10^4 $ &  $ 6.0\times10^4 $ &  $ 0.54 $ &  $ 0.43 $ \\ \hline
$ 215 $ &  $ 409\,019 $ &  $ 398\,000 $ &  $ 21\,000 $ &  $ 4.8\times10^4 $ &  $ 4.0\times10^4 $ &  $ 0.35 $ &  $ 0.29 $ \\ \hline
$ 233 $ &  $ 289\,557 $ &  $ 284\,000 $ &  $ 15\,000 $ &  $ 3.2\times10^4 $ &  $ 2.8\times10^4 $ &  $ 0.23 $ &  $ 0.20 $ \\ \hline
$ 254 $ &  $ 206\,096 $ &  $ 202\,000 $ &  $ 10\,000 $ &  $ 2.3\times10^4 $ &  $ 2.0\times10^4 $ &  $ 0.16 $ &  $ 0.14 $ \\ \hline
$ 276 $ &  $ 146\,653 $ &  $ 144\,000 $ &  $ 7000 $ &  $ 1.6\times10^4 $ &  $ 1.4\times10^4 $ &  $ 0.12 $ &  $ 0.10 $ \\ \hline
$ 300 $ &  $ 104\,516 $ &  $ 103\,000 $ &  $ 5000 $ &  $ 1.1\times10^4 $ &  $ 1.0\times10^4 $ &  $ 0.083 $ &  $ 0.073 $ \\ \hline
$ 326 $ &  $ 74\,371 $ &  $ 73\,000 $ &  $ 4000 $ &  $ 8.3\times10^3 $ &  $ 7.4\times10^3 $ &  $ 0.059 $ &  $ 0.053 $ \\ \hline
$ 354 $ &  $ 52\,871 $ &  $ 52\,100 $ &  $ 2900 $ &  $ 6.1\times10^3 $ &  $ 5.5\times10^3 $ &  $ 0.044 $ &  $ 0.039 $ \\ \hline
$ 385 $ &  $ 37\,630 $ &  $ 37\,100 $ &  $ 2200 $ &  $ 4.5\times10^3 $ &  $ 4.1\times10^3 $ &  $ 0.032 $ &  $ 0.030 $ \\ \hline
$ 419 $ &  $ 26\,878 $ &  $ 26\,300 $ &  $ 1600 $ &  $ 3.5\times10^3 $ &  $ 3.1\times10^3 $ &  $ 0.025 $ &  $ 0.022 $ \\ \hline
$ 455 $ &  $ 19\,035 $ &  $ 18\,700 $ &  $ 1200 $ &  $ 2.6\times10^3 $ &  $ 2.3\times10^3 $ &  $ 0.018 $ &  $ 0.017 $ \\ \hline
$ 495 $ &  $ 13\,578 $ &  $ 13\,200 $ &  $ 900 $ &  $ 2.0\times10^3 $ &  $ 1.7\times10^3 $ &  $ 0.014 $ &  $ 0.012 $ \\ \hline
$ 538 $ &  $ 9565 $ &  $ 9400 $ &  $ 700 $ &  $ 1.4\times10^3 $ &  $ 1.3\times10^3 $ &  $ 0.010 $ &  $ 0.0093 $ \\ \hline
$ 585 $ &  $ 6804 $ &  $ 6600 $ &  $ 500 $ &  $ 1.1\times10^3 $ &  $ 9.6\times10^2 $ &  $ 0.0080 $ &  $ 0.0069 $ \\ \hline
$ 635 $ &  $ 4754 $ &  $ 4600 $ &  $ 400 $ &  $ 8.0\times10^2 $ &  $ 7.1\times10^2 $ &  $ 0.0058 $ &  $ 0.0051 $ \\ \hline
$ 691 $ &  $ 3353 $ &  $ 3250 $ &  $ 280 $ &  $ 6.1\times10^2 $ &  $ 5.3\times10^2 $ &  $ 0.0044 $ &  $ 0.0038 $ \\ \hline
$ 751 $ &  $ 2297 $ &  $ 2240 $ &  $ 210 $ &  $ 4.3\times10^2 $ &  $ 3.9\times10^2 $ &  $ 0.0031 $ &  $ 0.0028 $ \\ \hline
$ 816 $ &  $ 1624 $ &  $ 1520 $ &  $ 150 $ &  $ 3.6\times10^2 $ &  $ 2.8\times10^2 $ &  $ 0.0026 $ &  $ 0.0020 $ \\ \hline
$ 887 $ &  $ 1093 $ &  $ 1020 $ &  $ 110 $ &  $ 2.6\times10^2 $ &  $ 2.0\times10^2 $ &  $ 0.0018 $ &  $ 0.0014 $ \\ \hline
$ 965 $ &  $ 754 $ &  $ 700 $ &  $ 80 $ &  $ 1.9\times10^2 $ &  $ 1.5\times10^2 $ &  $ 0.0014 $ &  $ 0.0011 $ \\ \hline
$ 1049 $ &  $ 517 $ &  $ 470 $ &  $ 60 $ &  $ 1.4\times10^2 $ &  $ 1.1\times10^2 $ &  $ 0.0010 $ &  $ 0.00078 $ \\ \hline
$ 1140 $ &  $ 367 $ &  $ 320 $ &  $ 40 $ &  $ 1.2\times10^2 $ &  $ 80 $ &  $ 0.00086 $ &  $ 0.00057 $ \\ \hline
$ 1239 $ &  $ 262 $ &  $ 215 $ &  $ 29 $ &  $ 1.0\times10^2 $ &  $ 60 $ &  $ 0.00073 $ &  $ 0.00043 $ \\ \hline
$ 1347 $ &  $ 166 $ &  $ 143 $ &  $ 21 $ &  $ 64 $ &  $ 44 $ &  $ 0.00046 $ &  $ 0.00032 $ \\ \hline
$ 1465 $ &  $ 113 $ &  $ 95 $ &  $ 15 $ &  $ 49 $ &  $ 33 $ &  $ 0.00035 $ &  $ 0.00024 $ \\ \hline
$ 1592 $ &  $ 77 $ &  $ 63 $ &  $ 11 $ &  $ 38 $ &  $ 26 $ &  $ 0.00027 $ &  $ 0.00018 $ \\ \hline
$ 1731 $ &  $ 48 $ &  $ 41 $ &  $ 8 $ &  $ 25 $ &  $ 19 $ &  $ 0.00018 $ &  $ 0.00014 $ \\ \hline
$ 1882 $ &  $ 30 $ &  $ 27 $ &  $ 6 $ &  $ 18 $ &  $ 15 $ &  $ 0.00013 $ &  $ 0.00011 $ \\ \hline
$ 2046 $ &  $ 21 $ &  $ 18 $ &  $ 4 $ &  $ 15 $ &  $ 13 $ &  $ 0.00011 $ &  $ 9.0\times10^{-5} $ \\ \hline
$ 2224 $ &  $ 16 $ &  $ 11.4 $ &  $ 3.1 $ &  $ 14 $ &  $ 9.5 $ &  $ 0.00010 $ &  $ 6.8\times10^{-5} $ \\ \hline
$ 2418 $ &  $ 8 $ &  $ 7.4 $ &  $ 2.2 $ &  $ 8.6 $ &  $ 7.7 $ &  $ 6.2\times10^{-5} $ &  $ 5.5\times10^{-5} $ \\ \hline
$ 2628 $ &  $ 5 $ &  $ 4.7 $ &  $ 1.6 $ &  $ 6.9 $ &  $ 6.9 $ &  $ 5.0\times10^{-5} $ &  $ 5.0\times10^{-5} $ \\ \hline
$ 2857 $ &  $ 3 $ &  $ 3.0 $ &  $ 1.1 $ &  $ 5.6 $ &  $ 5.6 $ &  $ 4.1\times10^{-5} $ &  $ 4.1\times10^{-5} $ \\ \hline
$ 3106 $ &  $ 2 $ &  $ 1.9 $ &  $ 0.8 $ &  $ 5.0 $ &  $ 5.0 $ &  $ 3.6\times10^{-5} $ &  $ 3.6\times10^{-5} $ \\ \hline
$ 3377 $ &  $ 2 $ &  $ 1.2 $ &  $ 0.5 $ &  $ 5.3 $ &  $ 4.1 $ &  $ 3.8\times10^{-5} $ &  $ 2.9\times10^{-5} $ \\ \hline
$ 3671 $ &  $ 1 $ &  $ 0.8 $ &  $ 0.4 $ &  $ 4.2 $ &  $ 4.2 $ &  $ 3.1\times10^{-5} $ &  $ 3.1\times10^{-5} $ \\ \hline
$ 3990 $ &  $ 1 $ &  $ 0.47 $ &  $ 0.25 $ &  $ 4.4 $ &  $ 3.0 $ &  $ 3.2\times10^{-5} $ &  $ 2.2\times10^{-5} $ \\ \hline
$ 4338 $ &  $ 1 $ &  $ 0.29 $ &  $ 0.16 $ &  $ 4.5 $ &  $ 3.0 $ &  $ 3.2\times10^{-5} $ &  $ 2.2\times10^{-5} $ \\ \hline
$ 4716 $ &  $ 1 $ &  $ 0.18 $ &  $ 0.11 $ &  $ 4.6 $ &  $ 3.0 $ &  $ 3.3\times10^{-5} $ &  $ 2.2\times10^{-5} $ \\ \hline
$ 5127 $ &  $ 0 $ &  $ 0.11 $ &  $ 0.07 $ &  $ 3.0 $ &  $ 3.0 $ &  $ 2.2\times10^{-5} $ &  $ 2.2\times10^{-5} $ \\ \hline
\end{tabular}
\label{tab:modelIndepLimits_mu}
\end{table}
 
\clearpage
\printbibliography

\clearpage
 
\begin{flushleft}
{\Large The ATLAS Collaboration}

\bigskip

G.~Aad$^\textrm{\scriptsize 101}$,    
B.~Abbott$^\textrm{\scriptsize 128}$,    
D.C.~Abbott$^\textrm{\scriptsize 102}$,    
O.~Abdinov$^\textrm{\scriptsize 13,*}$,    
A.~Abed~Abud$^\textrm{\scriptsize 70a,70b}$,    
K.~Abeling$^\textrm{\scriptsize 53}$,    
D.K.~Abhayasinghe$^\textrm{\scriptsize 93}$,    
S.H.~Abidi$^\textrm{\scriptsize 167}$,    
O.S.~AbouZeid$^\textrm{\scriptsize 40}$,    
N.L.~Abraham$^\textrm{\scriptsize 156}$
% ...
\bigskip
\\

% ...
$^{13}$Institute of Physics, Azerbaijan Academy of Sciences, Baku; Azerbaijan.\\
% ...
$^{40}$Niels Bohr Institute, University of Copenhagen, Copenhagen; Denmark.\\
% ...
$^{53}$II. Physikalisches Institut, Georg-August-Universit\"{a}t G\"ottingen, G\"ottingen; Germany.\\
% ...
$^{70}$$^{(a)}$INFN Sezione di Pavia;$^{(b)}$Dipartimento di Fisica, Universit\`a di Pavia, Pavia; Italy.\\
% ...
$^{93}$Department of Physics, Royal Holloway University of London, Egham; United Kingdom.\\
% ...
$^{101}$CPPM, Aix-Marseille Universit\'e, CNRS/IN2P3, Marseille; France.\\
$^{102}$Department of Physics, University of Massachusetts, Amherst MA; United States of America.\\
% ...
$^{128}$Homer L. Dodge Department of Physics and Astronomy, University of Oklahoma, Norman OK; United States of America.\\
% ...
$^{156}$Department of Physics and Astronomy, University of Sussex, Brighton; United Kingdom.\\
% ...
$^{167}$Department of Physics, University of Toronto, Toronto ON; Canada.\\
% ...
$^{a}$ Also at CERN, Geneva; Switzerland.\\
$^{b}$ Also at CPPM, Aix-Marseille Universit\'e, CNRS/IN2P3, Marseille; France.\\
% ...
$^{*}$ Deceased

\end{flushleft}


\end{document}